\documentclass{article}%
\usepackage{amsmath}
\usepackage{graphicx}%
\usepackage{amsfonts}%
\usepackage{amssymb}

\begin{document}

\title{Information in statistical physics \thanks{\textit{Submitted to:} Studies in
History and Philosophy of Modern Physics}}
\author{Roger Balian\\Service de Physique Th\'{e}orique\\CEA/Saclay - DSM/SPhT\\F-91191 Gif sur Yvette Cedex, France\\balian@spht.saclay.cea.fr}
\date{}
\maketitle

\begin{abstract}
We review with a tutorial scope the information theory foundations of quantum
statistical physics. Only a small proportion of the variables that
characterize a system at the microscopic scale can be controlled, for both
practical and theoretical reasons, and a probabilistic description involving
the observers is required. The criterion of maximum von~Neumann entropy is
then used for making reasonable inferences. It means that no spurious
information is introduced besides the known data. Its outcomes can be given a
direct justification based on the principle of indifference of Laplace. We
introduce the concept of relevant entropy associated with some set of relevant
variables; it characterizes the information that is missing at the microscopic
level when only these variables are known. For equilibrium problems, the
relevant variables are the conserved ones, and the Second Law is recovered as
a second step of the inference process. For non-equilibrium problems, the
increase of the relevant entropy expresses an irretrievable loss of
information from the relevant variables towards the irrelevant ones. Two
examples illustrate the flexibility of the choice of relevant variables and
the multiplicity of the associated entropies: the thermodynamic entropy
(satisfying the Clausius--Duhem inequality) and the Boltzmann entropy
(satisfying the $H$-theorem). The identification of entropy with missing
information is also supported by the paradox of Maxwell's demon. Spin-echo
experiments show that irreversibility itself is not an absolute concept: use
of hidden information may overcome the arrow of time.

\end{abstract}

\noindent\textit{keywords}: quantum probabilities, inference, reduced
description, relevant \newline entropies, irreversibility paradox.\newpage

\section{Probabilities in theoretical physics}

The purpose of this contribution is to present an overview of the approach to
statistical mechanics based on information theory. The concept of information,
intimately connected with that of probability, gives indeed insight on
questions of statistical mechanics such as the meaning of irreversibility.
This concept was introduced in statistical physics by Brillouin (1956) and
Jaynes (1957) soon after its discovery by Shannon in 1948 (Shannon and Weaver,
1949). An immense literature has since then been published, ranging from
research articles to textbooks. The variety of topics that belong to this
field of science makes it impossible to give here a bibliography, and special
searches are necessary for deepening the understanding of one or another
aspect. For tutorial introductions, somewhat more detailed than the present
one, see R. Balian (1991-92; 2004).

The meaning of probabilities, on which the concept of information relies, has
long been a subject of controversy (Cox, 1946). They were introduced in the
theory of games as frequencies of occurrence, directly related to the counting
of configurations, and could be viewed in this framework as properties of the
system \textit{in itself}. This objective interpretation contrasts with the
so-called subjective interpretation, which was initiated by Bayes and Laplace,
and later on advocated by Jaynes (1957) and de Finetti (1974).

While the mathematicians who focus on the structure of the theory of
probabilities do not adopt a definite position on their objective or
subjective meaning, the physicists who apply them should worry about their
interpretation. Most physicists have reached a spontaneous, often implicit,
philosophical position inspired by their daily practice, which combines
subjectivism and materialism. During the XIXth century, science was generally
considered as a discovery of Laws of nature existing outside us; in such an
activity the observer can be disregarded once the exploration stage is
over.\ The development of microphysics during the XXth century has led the
physicists to lay more emphasis on the role of the observer. Theory is
generally regarded as the construction of a partly imperfect image of the
external world in our minds, which however becomes less and less blurred, more
and more faithful as science progresses. Mathematics provide the precise
language which allows us to build this image; among mathematics, probabilities
are the tool on which we rely to make quantitative predictions in a consistent
and rational way, starting from the available information, in spite of various
uncertainties. It is remarkable that the introduction of probabilities in our
most fundamental theories has not prevented physics to become more and more efficient.

The existence of barriers between the reality and us, which cannot be overcome
at least at the present stage of physics, has become manifest in two different
theoretical frameworks. On the one hand, the unification and the simplicity of
the Laws of microphysics have led to a considerable development of
\textit{statistical mechanics}. It has become clear that all properties of
materials at the macroscopic scale, whether mechanical, thermal or
electromagnetic, can \textit{in principle} be explained by starting from their
microscopic structure. Macroscopic theories are then reduced to a
phenomenological status. Even the Laws of thermodynamics have now lost their
status of fundamental science since they can be derived from microphysics.
However, describing fully a macroscopic object in terms of its microscopic
constituents would require to deal with an inaccessibly large number of
degrees of freedom. Neither can we control them experimentally nor even can we
handle them numerically. Our only issue is a probabilistic treatment, where
probabilities account for our lack of knowledge about, for instance, the
positions and velocities of the molecules in a classical gas.

On the other hand, microscopic systems are governed by \textit{quantum
mechanics}. There, physical quantities are mathematically represented as
recalled in section 2 by elements of a non-commutative algebra. This feature
implies, in particular, Heisenberg's inequality which expresses that two
non-commuting variables (such as the position and the momentum of a particle,
the three components of its angular momentum, the electromagnetic field and
the number of photons, or the electric and magnetic fields at the same point)
necessarily display statistical fluctuations: The values of such variables
cannot be specified simultaneously. Here we need probabilities not only for
practical reasons as in classical statistical mechanics, but because the
fundamental theory itself implies \textit{intrinsic fluctuations}. It is not
simply the \textit{values} of the physical variables which are incompletely
known, but it is the very \textit{concept} of physical quantities which, as a
matter of principle, makes them simultaneously inaccessible. Their
non-commutative nature forbids us to imagine that we might fully determine
them. It is impossible to assume that, underlying our probabilistic
description, the state of a microscopic system at a given time could be
characterized by the values of its full set of physical variables. This
impossibility, which contrasts with the use of probabilities in the
description of systems in classical statistical mechanics, is exemplified by
Bell's inequalities and by the GHZ paradox (Greenberger et al, 1990) that we
shall briefly review at the end of section 2.

In both cases, and a fortiori in quantum statistical mechanics, we are led to
always treat physical quantities as random. The probabilities that govern them
occur either for practical reasons because we cannot hope to describe in full
detail a macroscopic system, or for theoretical reasons because quantum
fluctuations prevent us from thinking that the whole set of physical
quantities needed to describe a system in the quantum formalism may take
well-defined values. Thus the probabilities of quantum statistical physics
cannot be regarded as properties of a system as such, but they characterize
the knowledge about this system of its observers in the considered conditions.
The probability law adequately describing the system depends on the
information available to the observer or on the number of variables that he
may control. Probabilities therefore appear as having a partly subjective
nature, on rather inter-subjective since two observers placed in the same
conditions will assign the same probability law to a system.

Moreover, such a probabilistic description does not refer to a single object.
Explicitly or implicitly we regard this object as belonging to a statistical
\textit{ensemble} of similar objects, all prepared under similar conditions
and characterized by the same set of given data. This ensemble may be real, if
we deal with predictions about repeated experiments, or gedanken, if we deal
with a single event.

Even though quantum mechanics requires the use of probabilities to describe
systems at the microscopic scale, and even though these probabilities
characterize the knowledge of observers about such systems, the theory also
displays \textit{objective features} that depend on the systems only. Indeed,
as we shall recall below, the microscopic equations of motion of an isolated
system are fully deterministic since the Hamiltonian operator is exactly
known. (For most properties of atoms, molecules and macroscopic materials,
only the very simple kinetic energy and electromagnetic interactions of the
elementary constituents, the nuclei and the electrons, are relevant.) Whereas
at each time the statistics of the physical quantities are accounted for by a
probability law, the \textit{evolution} of this law is governed at the
microscopic scale by the reversible equation (\ref{002}) below, which depends
only on the object studied. The probability distribution, which characterizes
our knowledge, is transferred from one time to another in a deterministic way
-- unless we drop information or unless part of the Hamiltonian is ill-known.

Other objective features can also emerge at the macroscopic scale. Owing to
the large number of constituents, the probability distribution for many
macroscopic variables can be sharply peaked. The uncertainties of the
observers and the subjective aspect of probabilities can then be disregarded,
and statistical predictions are replaced by objective assertions, which hold
even for a single object.

The first stage of an inference consists in assigning to the system a
probability distribution that accounts for the available data but is otherwise
unbiased. Then, predictions are derived from this distribution by means of
standard techniques of statistical physics. For instance, we may wish to
predict the two-particle correlation function in a simple liquid at
equilibrium, so as to understand how it scatters light; in this problem the
data which characterize the macroscopic equilibrium state are the energy and
particle number per unit volume. This correlation function can be deduced from
the density in phase, that is, the probability distribution of the $N$
particles of the liquid in the $6N$-dimensional phase space. However a
preliminary question should be solved: From the sole knowledge of the energy
and the particle number, how should we reasonably choose this probability distribution?

\section{The formalism of quantum (statistical) mechanics}

We shall work within the framework of quantum statistical mechanics (Thirring,
1981, 1983; Balian, 1989), which is conceptually somewhat simpler than
classical statistical mechanics. To avoid mathematical complications, we
consider finite systems only and assume that the thermodynamic limit for
extensive systems is taken in the end. The discreteness of spectra is then a
feature which allows to by-pass some difficulties arising in classical
statistical mechanics from the continuity of the variables. For instance, the
classical limit of quantum statistical mechanics generates the suitable
measure in the $6N$-dimensional phase space; this measure includes a factor
$\frac{1}{N!}$ issued from the Pauli principle for indistinguishable
particles, which ensures the extensivity of entropy and solves the Gibbs
paradox. These facts can be explained in the framework of classical
statistical mechanics (van Kampen, 1984), but less automatically. Obtaining
the Third Law of thermodynamics as a consequence of statistical mechanics also
requires quantization, since the behaviour at low temperatures of the entropy
of a macroscopic system is governed by the low energy behaviour of its level density.

The mathematical description of a physical system in quantum mechanics
involves a Hilbert space, and the physical quantities are represented by the
Hermitean operators $\hat{A}$ in this space. As indicated in the introduction,
these operators, termed \textit{observables}, play the r\^{o}le of random
variables but constitute a non-commutative algebra. For instance, the three
components of the spin $\frac{1}{2}$ of a fermion are described in a
$2$-dimensional Hilbert space by the Pauli operators $\hat{\sigma}_{x}$,
$\hat{\sigma}_{y}$, $\hat{\sigma}_{z}$, which are characterized by the algebra
$\hat{\sigma}_{i}\hat{\sigma}_{j}=i\sum_{k}\varepsilon_{ijk}\hat{\sigma}_{k}$;
for a particle on a line, the algebra is generated by the position and
momentum operators $\hat{x}$ and $\hat{p}$ in the Hilbert space of
wavefunctions, with $\left[  \hat{x},\hat{p}\right]  =i\hbar$. The specific
features of quantum physics, compared to the standard probability theory or to
the classical statistical mechanics, lie in this non-commutation of the observables.

In quantum mechanics, the \textquotedblleft state of a
system\textquotedblright, whatever formalism is used to characterize it
(wavefunction, state vector, density matrix, etc), is an irreducibly
probabilistic concept. As stressed in the introduction, one cannot imagine the
existence of an underlying, purely objective description in terms of the
values of all the physical quantities attached to the system. We are led to
adhere to the Bayesian or Laplacian subjective conception of probabilities,
and to regard the \textquotedblleft state\textquotedblright\ as the best
possible description of the system, which allows us to make any possible
probabilistic prediction about it -- or rather about the statistical ensemble
to which it belongs.

A set of probabilities is equivalent to the collection of expectation values
that they allow us to evaluate. A \textit{quantum state} is thus characterized
by the \textit{correspondence} $\hat{A}\mapsto\left\langle \hat{A}%
\right\rangle \equiv A$ which associates with \textit{any observable} $\hat
{A}$ \textit{its expectation value} $A$ in the considered situation. This
correspondence has a few natural properties. The hermiticity of $\hat{A} $
entails that $A$ is real (quantum mechanics involves complex numbers, but
physical predictions yield real numbers). For the unit operator, $\left\langle
\hat{I}\right\rangle =1$. The correspondence is linear; in fact, the linearity
$\left\langle \hat{A}+\hat{B}\right\rangle =\left\langle \hat{A}\right\rangle
+\left\langle \hat{B}\right\rangle $ for any pair of \textit{commuting}
observables $\hat{A}$ and $\hat{B}$ is sufficient to imply linearity for the
whole set of observables (provided the dimension of the Hilbert space is
larger than $2$: Gleason's theorem). Finally the statistical fluctuation of a
physical quantity is expressed as $\left\langle \hat{A}^{2}\right\rangle
-\left\langle \hat{A}\right\rangle ^{2}$ in terms of the expectation values
(\ref{001}). Imposing that $\left\langle \hat{A}^{2}\right\rangle $ is
non-negative for any Hermitean $\hat{A}$ expresses that a variance cannot be negative.

For finite systems these properties are implemented through the existence of a
\textit{density operator} $\hat{D}$ in Hilbert space which represents the
state. The above correspondence from observables to their expectation values
is expressed as
\begin{equation}
\hat{A}\mapsto\left\langle \hat{A}\right\rangle \equiv A=\operatorname{Tr}%
\hat{A}{\hat{D}}\ ,\tag{1}\label{001}%
\end{equation}
where $\hat{D}$ is a Hermitean positive operator with unit trace. The density
operator thus gathers our whole probabilistic information about the full set
of observables of the system. It plays with respect to observables the same
r\^{o}le as a usual probability distribution does with respect to random
variables. A wavefunction or a state vector $\left\vert \Psi\right\rangle $
appears as nothing but a special case of density operator; in this case
$\hat{D}=\left\vert \Psi\right\rangle \left\langle \Psi\right\vert $ reduces
to a projection on the considered pure state and the expectation values
(\ref{001}) read $A=\left\langle \Psi\right\vert \hat{A}\left\vert
\Psi\right\rangle $. Note that, due to the irreducibly probabilistic nature of
the quantum theory, there is no conceptual difference between
\textquotedblleft quantum mechanics\textquotedblright\ and \textquotedblleft
quantum statistical mechanics\textquotedblright, since a wavefunction is
nothing but a means for evaluating probabilities.

For time-dependent problems, we need in order to make predictions to express
how this information (\ref{001}), given at some time, is transformed at later
times. This is expressed in the Schr\"{o}dinger picture by letting $\hat{D}$
depend on time while the observables $\hat{A}$ remain fixed. For an isolated
system, with given Hamiltonian $\hat{H}$, the evolution of $\hat{D}\left(
t\right)  $ is governed by the \textit{Liouville--von~Neumann equation} of
motion:
\begin{equation}
i\hbar\frac{d{\hat{D}}}{dt}=\left[  {\hat{H}},{\hat{D}}\right]  \text{\ .}%
\tag{2}\label{002}%
\end{equation}

Whereas the definition of quantum states involves both the system and the
observers and has a subjective aspect because the density operator appears as
a tool to make consistent predictions from the available information, the
evolution refers to the system only. This purely objective nature of the
evolution is made clear in the alternative Heisenberg picture, where the
density operator $\hat{D}$ remains fixed while the observables change in time
according to Heisenberg's equation of motion
\begin{equation}
i\hbar\frac{d\hat{A}}{dt}=\left[  \hat{A},\hat{H}\right]  \text{ ,}%
\tag{3}\label{003}%
\end{equation}
which is deterministic and reversible. Equations (\ref{002}) and (\ref{003})
are equivalent as regards the evolution of the expectation values (\ref{001}).
Heisenberg's equation (\ref{003}) exhibits better the fact that the dynamics
is a property of the considered system, since it deals with the observables
which represent mathematically the physical quantities belonging to this
system, independently of observation. In fact, eq. (\ref{003}) simply
expresses that the operator \textit{algebra remains unchanged} in time while
the various observables $\hat{A}$ evolve; this property is equivalent to the
unitary of the evolution, with the Hamiltonian as generator of the unitary
motion. As another advantage, Heisenberg's picture allows us to evaluate
correlations between observables of the same system taken at different times.
However, because we focus below on information, we shall start from
Schr\"{o}dinger's picture rather than Heisenberg's. When expressed by means of
(\ref{002}) instead of (\ref{003}), the evolution takes a subjective aspect,
since (\ref{002}) describes the transfer of information from some observables
to other ones generated by a completely known dynamics. We can then modify eq.
(\ref{002}) so as to account for losses of information that may take place
during a dynamical process.

In the \textit{classical limit}, observables $\hat{A}$\ are replaced by
commuting random variables, which are functions of the positions and momenta
of the $N$\ particles. Density operators $\hat{D}$\ are replaced by
probability densities $D$\ in the $6N$-dimensional phase space, and the trace
in (\ref{001}) by an integration over this space. The evolution of $D$ is
governed by the Liouville equation.

To conclude this section, we show how the \textit{irreducibly probabilistic}
nature of quantum states follows from the non-commutation of the observables.
Consider first two physical quantities represented by two observables $\hat
{A}$ and $\hat{B}$ which do not commute, $\left[  \hat{A},\hat{B}\right]
=2i\hat{C}$. Since for any $\lambda$ the operator $\left(  \hat{A}%
+i\lambda\hat{B}\right)  \left(  \hat{A}-i\lambda\hat{B}\right)  $ can be
regarded as the square of a Hermitean operator (as obvious by
diagonalization), its expectation value is non-negative, which implies
$\left\langle \hat{A}^{2}\right\rangle \left\langle \hat{B}^{2}\right\rangle
\geq\left\langle \hat{C}\right\rangle ^{2}$. This Heisenberg's inequality sets
a lower bound to statistical fluctuations (for instance, $2\hat{C}=\hbar$ for
$\hat{A}=\hat{x}$, $\hat{B}=\hat{p}$ yields $\Delta\hat{x}\Delta\hat{p}%
\geq\hbar/2$). Accordingly, non-commuting physical quantities are
\textit{incompatible}: They cannot be measured nor even specified simultaneously.

Bell's inequalities and the GHZ paradox (Greenberger et al, 1990) also arise
from non-commuta\-tion. Consider, for instance, three spins
\textbf{\mbox{\boldmath{$\hat{\sigma}$}}}$^{\left(  m\right)  }$, $m=1$, $2$,
$3$; define the observables $\hat{A}^{\left(  1\right)  }\equiv\hat{\sigma
}_{x}^{\left(  1\right)  }$, $\hat{B}^{\left(  1\right)  }\equiv\hat{\sigma
}_{z}^{\left(  2\right)  }\hat{\sigma}_{z}^{\left(  3\right)  }$, $\hat
{C}^{\left(  1\right)  }\equiv\hat{A}^{\left(  1\right)  }\hat{B}^{\left(
1\right)  }$, and $\hat{A}^{\left(  m\right)  }$, $\hat{B}^{\left(  m\right)
}$, $\hat{C}^{\left(  m\right)  }$ by cyclic permutation. All of them have
$+1$ and $-1$ as eigenvalues, the $\hat{A}$'s commute with one another as well
as the $\hat{B}$'s and $A^{\left(  m\right)  }$ commutes with $\hat
{B}^{\left(  m\right)  }$. For $m\neq n$, $\hat{A}^{\left(  m\right)  }$ and
$\hat{B}^{\left(  n\right)  }$ anticommute, so that the three observables
$\hat{C}^{\left(  n\right)  }$ commute with one another. Hence, we can imagine
that the $3$-spin system is prepared in a pure state $\left\vert
\Psi\right\rangle $ which is the common eigenvector of $\hat{C}^{\left(
1\right)  }$, $\hat{C}^{\left(  2\right)  }$ and $\hat{C}^{\left(  3\right)
}$ characterized by their eigenvalues $c^{\left(  1\right)  }$, $c^{\left(
2\right)  }$, $c^{\left(  3\right)  }$ all equal to $+1$. In this state, the
fact that $\hat{C}^{\left(  1\right)  }\equiv\hat{A}^{\left(  1\right)  }%
\hat{B}^{\left(  1\right)  }$ takes the value $c^{\left(  1\right)  }=+1$
implies that $a^{\left(  1\right)  }=b^{\left(  1\right)  }$; more precisely,
if we were to measure $\hat{A}^{\left(  1\right)  }$ and $\hat{B}^{\left(
1\right)  }$ simultaneously (they commute), we would find for them either the
values $a^{\left(  1\right)  }=b^{\left(  1\right)  }=+1$ or the values
$a^{\left(  1\right)  }=b^{\left(  1\right)  }=-1$ (with equal probabilities
since the expectation value of $\hat{A}^{\left(  1\right)  }$ or $\hat
{B}^{\left(  1\right)  }$ vanishes). Likewise we can assert the complete
correlations $a^{\left(  2\right)  }=b^{\left(  2\right)  }$ and $a^{\left(
3\right)  }=b^{\left(  3\right)  }$. However the above definitions and
algebraic relations imply the operator identities $\hat{B}^{\left(  1\right)
}\hat{B}^{\left(  2\right)  }\hat{B}^{\left(  3\right)  }\equiv\hat{I}$ and
$\hat{C}^{\left(  1\right)  }\hat{C}^{\left(  2\right)  }\hat{C}^{\left(
3\right)  }\equiv-\hat{A}^{\left(  1\right)  }\hat{A}^{\left(  2\right)  }%
\hat{A}^{\left(  3\right)  }$. Hence, although the three statements
$a^{\left(  1\right)  }=b^{\left(  1\right)  }$, $a^{\left(  2\right)
}=b^{\left(  2\right)  }$ and $a^{\left(  3\right)  }=b^{\left(  3\right)  }$
(in the above sense) are \textit{separately true}, they cannot be true
\textit{together}: Since the product $\hat{A}^{\left(  1\right)  }\hat
{A}^{\left(  2\right)  }\hat{A}^{\left(  3\right)  }$ takes the value
$-c^{\left(  1\right)  }c^{\left(  2\right)  }c^{\left(  3\right)  }=-1$ in
the considered state $\left\vert \Psi\right\rangle $, the simultaneous
measurement of $\hat{A}^{\left(  1\right)  }$, $\hat{A}^{\left(  2\right)  }$
and $\hat{A}^{\left(  3\right)  }$ (which commute) is expected to yield values
$a^{\left(  1\right)  }$, $a^{\left(  2\right)  }$ and $a^{\left(  3\right)
}$ equal to $+1$ or $-1$, but necessarily with a product $a^{\left(  1\right)
}a^{\left(  2\right)  }a^{\left(  3\right)  }=-1$, in contradiction to the
naive prediction $a^{\left(  1\right)  }a^{\left(  2\right)  }a^{\left(
3\right)  }=+1$ which would result from $a^{\left(  1\right)  }=b^{\left(
1\right)  }$, $a^{\left(  2\right)  }=b^{\left(  2\right)  }$, $a^{\left(
3\right)  }=b^{\left(  3\right)  }$, $b^{\left(  1\right)  }b^{\left(
2\right)  }b^{\left(  3\right)  }=+1$. This has been experimentally confirmed.
Thus, everyday's logics is violated.

The only issue is to regard the \textquotedblleft state\textquotedblright\ not
as a property of the system alone, but as a probabilistic means for
prediction. The current expression \textquotedblleft the state of the
system\textquotedblright\ that we used (and that we may still use for
convenience below) is improper; it is meant as \textquotedblleft the
probability distribution (for non-commuting physical quantities) which allows
us to predict the properties of the statistical ensemble in which the system
is embedded\textquotedblright. Indeed, we can safely predict that a
measurement of $\hat{A}^{\left(  1\right)  }$ and $\hat{B}^{\left(  1\right)
}$ will give $a^{\left(  1\right)  }=b^{\left(  1\right)  }$ for any element
of the ensemble, and likewise for $\hat{A}^{\left(  2\right)  }$ and $\hat
{B}^{\left(  2\right)  }$. However, such measurements must be performed with
\textit{different apparatuses} on \textit{different samples} (described by the
same state vector $\left\vert \Psi\right\rangle $), because $\hat{A}^{\left(
1\right)  }$ and $\hat{B}^{\left(  2\right)  }$ do not commute and hence
cannot be measured nor even specified simultaneously. We cannot find in the
ensemble any system for which $\hat{A}^{\left(  1\right)  }$ takes with
certainty, say, the value $a^{\left(  1\right)  }=+1$ and $\hat{B}^{\left(
2\right)  }$ the value $b^{\left(  2\right)  }=+1$: The assertion that
\textit{both} $a^{\left(  1\right)  }=b^{\left(  1\right)  }$ \textit{and}
$a^{\left(  2\right)  }=b^{\left(  2\right)  }$ is meaningless. The
contradiction existing between the equalities $a^{\left(  1\right)
}=b^{\left(  1\right)  }$, $a^{\left(  2\right)  }=b^{\left(  2\right)  }$,
$a^{\left(  3\right)  }=b^{\left(  3\right)  }$, $b^{\left(  1\right)
}b^{\left(  2\right)  }b^{\left(  3\right)  }=1$ and $a^{\left(  1\right)
}a^{\left(  2\right)  }a^{\left(  3\right)  }=-1$ implies that we should
\textit{not} regard the two correlations $a^{\left(  1\right)  }=b^{\left(
1\right)  }$ and $a^{\left(  2\right)  }=b^{\left(  2\right)  }$ as
\textit{intrinsic properties} of the system, but rather consider \textit{each
one as an exact prediction about a specific measurement}, namely that of
$\hat{A}^{\left(  1\right)  }$ and $\hat{B}^{\left(  1\right)  }$ in one case,
that of $\hat{A}^{\left(  2\right)  }$ and $\hat{B}^{\left(  2\right)  }$ in
the other case, the measurements of $\hat{A}^{\left(  1\right)  }$ and
$\hat{B}^{\left(  2\right)  }$ being incompatible. The state vector
$\left\vert \Psi\right\rangle $ itself, which synthesizes all such
information, does not describe intrinsic properties of the systems belonging
to the considered ensemble, but rather tells us how they would behave in some
experiment or another. Each assertion such as $a^{\left(  1\right)
}=b^{\left(  1\right)  }$ (which to be tested would require an interaction of
a system with an apparatus measuring $\hat{A}^{\left(  1\right)  }$ and
$\hat{B}^{\left(  1\right)  }$), is true \textit{only in a given context},
which should be specified even if the measurement is not actually performed.
The interpretation of quantum mechanics is thus tightly connected with
measurement theory. The type of correlations that it involves cannot be
accounted for in terms of hidden variables.

\section{The measure of uncertainty}

The knowledge embodied in a density operator $\hat{D}$ is always
probabilistic, since $\hat{D}$ provides us only with the expectation values
(\ref{001}). At least some among the variances evaluated from (\ref{001}) must
be finite, due to non-commutation, because there exists no common eigenvector
for the whole set of observables. It is therefore natural to wonder whether
one density operator is more informative than another (Thirring, 1983; Balian,
1991). To this aim we associate with each $\hat{D}$ a number, the
\textit{von~Neumann entropy}
\begin{equation}
S_{\mathrm{vN}}\left(  {\hat{D}}\right)  =-\operatorname{Tr}{\hat{D}\ }%
\ln{\hat{D}}\text{{\ ,}}\tag{4}\label{004}%
\end{equation}
which measures in dimensionless units our uncertainty when $\hat{D}$
summarizes our statistical knowledge on the system. This quantity is the
quantum analogue of Shannon's entropy $-\sum_{m}p_{m}\ \ln p_{m}$; the latter
number measures the amount of \textit{information which is missing} on average
when we wait for the reception of some message belonging to a set $\left\{
m\right\}  $, if each message $m$ is expected to occur with a probability
$p_{m}$ (Shannon and Weaver, 1949). However, here, there will be no reception
of a message since we shall never observe fully the system at the microscopic
scale. The quantity $S\left(  \hat{D}\right)  $ qualifies merely our present
knowledge through $\hat{D}$, without reference to future observations which
can change our information.

We may also interpret $S\left(  \hat{D}\right)  $ as a measure of the
\textit{disorder} that we are facing when our information is characterized by
$\hat{D}$. This identification of the concept of disorder with a lack of
knowledge may seem questionable, since disorder seems an objective property of
the system whereas information is attached to the observer. However, even in
everyday's life, we can recognize that disorder is a relative property. My
desk may look quite disordered to a visitor, but it is \textit{for me} in
perfect order, if I am able to retrieve immediately any one of the many files
that lie scattered on it. Likewise, a pack of cards fully mixed by a conjurer
is disordered for the audience, but not for him, if he has kept knowledge of
the ordering of the cards. Information theory allows us to make quantitative
Maxwell's remark about the concept of order: ``Confusion, like the correlative
term order, is not a property of material things in themselves, but only in
relation to the mind who perceives them''.

Historically, Boltzmann was the first to identify, in many successive works,
the entropy of thermodynamics with a functional of the distribution in phase
space of classical statistical mechanics -- even though the concept of
probability was not yet fully settled. Von~Neumann extended this idea to
quantum physics by introducing the definition (\ref{004}). The existence of
these expressions inspired Shannon (1949) when he created the theory of
communication, which is based on the possibility of assigning numbers to
amounts of information. In this context, the subjective nature of entropy is
manifest. From the fact that it measures the missing information associated
with the probability distribution for the considered set of messages, it
appeared that it was a general concept complementing probability. With this
new interpretation as lack of information at the microscopic scale, the
concept has returned to statistical physics (Brillouin, 1956; Jaynes, 1957),
throwing new lights on entropy.

The von~Neumann entropy is characterized by many properties which confirm its
interpretation as a measure of \textit{uncertainty}. It is \textit{additive}
for uncorrelated systems, \textit{subadditive} for correlated ones (which
means that suppressing correlations raises the uncertainty), \textit{concave}
(which means that putting together two different statistical ensembles for the
same system produces a mixed ensemble which has a larger uncertainty than the
average of the uncertainties of the original ensembles). The \textit{maximum}
$\ln W$ of (\ref{004}) is obtained, for a finite Hilbert space with dimension
$W$ when $\hat{D}=\hat{I}/W$. Its \textit{minimum} $0$ is obtained for pure
states $\hat{D}$, which are the least uncertain states that quantum mechanics
allows, although they are still probabilistic.

\section{Maximum entropy criterion}

The availability of the density operator $\hat{D}$ at a given time $t_{0}$
allows us to make any statistical prediction on the considered system (or more
precisely on a statistical ensemble to which it belongs), either at the same
time through (\ref{001}), or at later times through (\ref{002}). However, a
preliminary problem arises. During the preparation of the system, before the
time $t_{0}$, only a small set of data are controlled. Let us denote as
$\hat{A}_{i}$ the observables that are controlled, and as $A_{i}$ their
expectation values for the considered set of repeated experiments. From this
\textit{partial information} we wish to \textit{infer} other quantities. In
other words, we wish to assign to the system a density operator $\hat{D}$, by
relying on the sole knowledge of the set
\begin{equation}
A_{i}=\operatorname{Tr}\hat{A}_{i}\hat{D}\text{\ .}\tag{5}\label{005}%
\end{equation}

The maximum entropy criterion consists in selecting, among all the density
operators subject to the \textit{constraints} (\ref{005}), the one, $\hat
{D}_{\mathrm{R}}$, which renders the \textit{von~Neumann entropy} (\ref{004})
\textit{maximum} (Jaynes, 1957). An intuitive justification is the following:
for any other $\hat{D}$ compatible with (\ref{005}), we have by construction
$S_{\mathrm{vN}}\left(  \hat{D}\right)  <S_{\mathrm{vN}}\left(  \hat
{D}_{\mathrm{R}}\right)  $. The choice of $\hat{D}_{\mathrm{R}}$ thus ensures
that our description involves \textit{no more information} than the minimum
needed to account for the only available information (\ref{005}). The
difference $S\left(  \hat{D}_{\mathrm{R}}\right)  -S\left(  \hat{D}\right)  $
measures some extra information included in $\hat{D}$, but not in $\hat
{D}_{\mathrm{R}}$, and hence, not in the only known expectation values $A_{i}%
$. Selecting $\hat{D}_{\mathrm{R}}$ rather than any other $\hat{D}$ which
satisfies (\ref{005}) is therefore the \textit{least biased choice}, the one
that allows the most reasonable predictions drawn from the known set $A_{i}$
about other arbitrary quantities (\ref{001}).

We may also interpret the maximum entropy criterion as the choice of the
\textit{most uncertain} state among those which satisfy the constraints
(\ref{005}) imposed by the data $A_{i}$, or equivalently of the \textit{most
disordered} one. This criterion may be regarded as a generalization of
Laplace's \textit{principle of insufficient reason} (or Bayes's principle),
which states that equal probabilities should be assigned to all possible
events in case nothing is known about the occurrence of these events. Indeed,
if the Hilbert space associated with the system has a finite dimension $W$,
the entropy is largest when $\hat{D}$ is proportional to the unit matrix
$\hat{I}$, so that the criterion yields $\hat{D}_{\mathrm{R}}=\hat{I}/W$,
describing equiprobability, in case there are no constraints (\ref{005}) on
expectation values.

This use of the lack of information as a tool for statistical inference has a
weakness: it relies on the assumption that $S_{\mathrm{vN}}$ is the adequate
measure of bias in statistical physics. Direct justifications of the outcome
of the maximum entropy criterion, based on some requirements of consistency of
the inference procedure, have therefore been given. They are discussed in
detail by Uffink (1995).

An alternative direct justification relies on the introduction of a
\textquotedblleft supersystem\textquotedblright, made of a large number
$\mathcal{N}$ of mental copies $\alpha=1,2,\ldots\mathcal{N}$ of the system in
hand. The statistical data are the same for all individual systems $\alpha$,
i.e., the observables $\hat{A}_{i,\alpha}\ (\alpha=1,2,\ldots\mathcal{N})$
have the same \textit{expectation} value $A_{i}$. This value is identified
with the value of the \textit{average} observable $\mathcal{N}^{-1}%
\sum_{\alpha}\hat{A}_{i,\alpha}$ pertaining to the supersystem, which has weak
statistical fluctuations (at most in $\mathcal{N}^{-1/2}$, due to
non-commutation of the observables $\hat{A}_{i}$). Such an identification
throws a bridge between the subjective and frequencial interpretations of
probabilities. The maximum entropy criterion then arises merely as a
consequence of the principle of insufficient reason applied to the
supersystem. This idea was introduced by Gibbs in the case where the only
datum (\ref{005}) is the energy: he showed that, for large $\mathcal{N}$, each
system $\alpha$ is in canonical equilibrium if a microcanonical distribution
is assigned to the supersystem. The extension to quantum statistical
mechanics, with non-commuting observables $\hat{A}_{i}$, was worked out by
Balian and Balazs (1987).

\section{Generalized Gibbsian distributions and relevant entropy}

For any given set of \textit{relevant observables} $\hat{A}_{i}$, the
expectation values $A_{i}$ of which are known, the maximum of the von~Neumann
entropy (\ref{004}) under the constraints (\ref{005}) is readily found by
introducing Lagrangian multipliers, $\gamma_{i}$ associated with each equation
(\ref{005}) and $\Psi$ associated with the normalization of $\hat{D}$. Its
value, $S_{\mathrm{vN}}\left(  \hat{D}_{\mathrm{R}}\right)  $, is reached for
a density operator of the form
\begin{equation}
\hat{D}_{\mathrm{R}}=\exp\left[  -\Psi-\sum_{i}\gamma_{i}\ \hat{A}_{i}\right]
\text{\ ,}\tag{6}\label{006}%
\end{equation}
where the multipliers are determined by
\begin{equation}
\operatorname{Tr}\hat{D}_{\mathrm{R}}\ \hat{A}_{i}=A_{i}\text{\quad,\quad
}\operatorname{Tr}\hat{D}_{\mathrm{R}}=1\text{\ .}\tag{7}\label{007}%
\end{equation}
This least biased density operator has an exponential form, which generalizes
the usual Gibbs distributions and which arises from the logarithm in the
definition of the von~Neumann entropy. The concavity of the von~Neumann
entropy ensures the unicity of $\hat{D}_{\mathrm{R}}$.

The equations (\ref{007}) can conveniently be written by introducing a
generalized thermodynamic potential $\Psi\left(  \gamma_{i}\right)  $, defined
as function of the other multipliers $\gamma_{i}$ by
\begin{equation}
\Psi\left(  \gamma_{i}\right)  \equiv\ln\operatorname{Tr}\exp\left[  -\sum
_{i}\gamma_{i}\hat{A}_{i}\right]  \text{\ .}\tag{8}\label{008}%
\end{equation}
The relations between the data $A_{i}$ and the multipliers $\gamma_{i}$ are
then implemented as
\begin{equation}
\frac{\partial\Psi}{\partial\gamma_{i}}=-A_{i}\text{\ .}\tag{9}\label{009}%
\end{equation}

The corresponding entropy $S_{\mathrm{vN}}\left(  \hat{D}_{\mathrm{R}}\right)
\equiv S_{\mathrm{R}}\left(  A_{i}\right)  $ is a function of the variables
$A_{i}$ (or equivalently $\gamma_{i}$), found from (\ref{004}), (\ref{005})
and (\ref{006}) as
\begin{equation}
S_{\mathrm{R}}\left(  A_{i}\right)  =\Psi+\sum_{i}\gamma_{i}A_{i}%
\text{\ .}\tag{10}\label{010}%
\end{equation}
We term it the \textit{relevant entropy} associated with the set $\hat{A}_{i}
$ of relevant observables selected in the considered situation. By
construction, it measures the amount of information which is missing
\textit{when only the data} $A_{i}$ \textit{are available}. Eq. (\ref{010})
exhibits $S_{\mathrm{R}}\left(  A_{i}\right)  $ as the Legendre transform of
$\Psi\left(  \gamma_{i}\right)  $. The relations (\ref{009}) between the
variables $\gamma_{i}$ and the variables $A_{i}$ can therefore be inverted as
\begin{equation}
\frac{\partial S_{\mathrm{R}}}{\partial A_{i}}=\gamma_{i}\text{\ .}%
\tag{11}\label{011}%
\end{equation}

The following examples will show that the concept of relevant entropy
encompasses various types of entropies, introduced in different contexts.

\section{Equilibrium thermodynamics}

As stressed by Callen (1975) the Laws of thermodynamics do \textit{not} deal
with \textit{dynamics}, but with the comparison of \textit{the initial and the
final equilibrium} states of a system. At the macroscopic level, an
equilibrium state is characterized by the values $A_{i}$ of a set of
\textit{extensive conservative} variables, such as energies or particle
numbers. In fact, the thermodynamic laws also hold for metastable states which
behave as stable states for small variations of the parameters. In such cases,
the thermodynamic variables $A_{i}$ include the nearly conserved quantities,
which strictly speaking obey no conservation law but are conserved over the
time scale of the considered process. This occurs, for instance, below a
transition point for an \textit{order parameter}, or when some a priori
allowed chemical reaction does not take place under the prevailing conditions.
The concept of equilibrium is thus relative: it depends on the efficiency of
the interactions and many equilibria are not truly stable, but metastable over
very long times. (As an extreme example, we may note that nuclear forces are
effective only within each nucleus of a material; although strong, they are
inhibited between one nucleus and another by the Coulomb repulsion. Nuclear
equilibrium, which would imply that nuclear reactions have transformed both
the light and the heavy nuclei through fusion or fission into iron nuclei, is
never reached.)

Callen's statement of the \textit{Second Law} refers to an \textit{isolated},
\textit{compound} system, in which we include the possible sources of heat or
work. Each subsystem is a homogeneous piece of extensive matter. In the data
$A_{i}$ that characterize the initial and the final state of the whole system,
the index $i$ denotes both the subsystem and the nature of the variable
(energy, number of particles of each kind, volume). Various exchanges of such
quantities may take place between the subsystems under the effect of
interactions. In the initial state these exchanges are prohibited by some
constraints, which express that the interactions between subsystems are
dropped, so that the variables $A_{i}$ can be specified independently. Later
on, some interactions are restored so that some constraints are released. We
wish to know the values of the variables $A_{i}$ in the final equilibrium
state thus reached. The answer is given by the Second Law, which postulates
the existence of the \textit{thermodynamic entropy} $S_{\mathrm{th}}$, a
function of the variables $A_{i}$ which has the following properties (Callen,
1975). It is \textit{additive}, being a sum of contributions of all
subsystems; for each subsystem, it is \textit{extensive} and \textit{concave}.
The final values of the variables $A_{i}$ are then found by looking for the
maximum of $S_{\mathrm{th}}$, under constraints imposed by the initial data
through the conservation laws and by possible inhibitions that forbid some
exchanges between the subsystems.

This formulation of the foundations of thermodynamics is directly related to
the approach to statistical mechanics of sections 4 and 5, based on
information theory. It suffices to identify the thermodynamic extensive
variables $A_{i}$ with the expectation values (\ref{005}) of the conserved
observables $\hat{A}_{i}$. More precisely, the $\hat{A}_{i}$'s are the
operators that commute with the Hamiltonian of the overall compound system
when the interactions between the subsystems are dropped. The expectation
values of the energies or of the particle numbers of the subsystems can then
be frozen. The macroscopic equilibrium state, characterized by the data
$A_{i}$, is identified microscopically with the most disordered state
(\ref{006}) accounting for these data. This surmise can be justified
statistically: if no other variable than these conserved quantities $A_{i}$ is
controlled, reasonable predictions on any other quantity\ should rely on the
least biased choice (\ref{006}). Moreover the variances of all the macroscopic
variables are then weak (of relative order $N^{-1}$, when $N$ is the particle
number); hence predictions about them can be made nearly with certainty, as we
observe macroscopically, and they are not affected by a coarse-graining of
(\ref{006}). The assignment of the density operator (\ref{006}) to an
equilibrium state can also be justified dynamically through non-equilibrium
statistical mechanics, by showing that the Hamiltonian evolution (\ref{002}),
while keeping the variables $A_{i}$ constant, leads to the coarse-grained
density operator with maximum disorder, as measured by the von~Neumann entropy.

We thus identify the macroscopic \textit{thermodynamic entropy}
$S_{\mathrm{th}}\left(  A_{i}\right)  $ with the microscopic \textit{relevant
entropy} $S_{\mathrm{R}}\left(  A_{i}\right)  $, which was defined as the
maximum of the von~Neumann entropy $S_{\mathrm{vN}}\left(  \hat{D}\right)  $
under the constraints (\ref{005}) on the thermodynamic variables $A_{i}$,
frozen in the initial state. The properties of $S_{\mathrm{th}}$ postulated in
thermodynamics are thereby found as consequences of statistical mechanics: its
additivity arises from the factorization of the density operator (\ref{006})
into contributions associated with each subsystem; its concavity results from
the concavity of $S_{\mathrm{vN}}$; its extensivity can be proved for a wide
class of interparticle interactions including the realistic interactions
(Ruelle, 1969; Lieb, 1976; Thirring,1983). The explicit form of
$S_{\mathrm{th}}\left(  A_{i}\right)  $ can also be derived from microphysics,
while only its existence is asserted in thermodynamics.

When applied to the final state, the maximum entropy criterion of statistical
mechanics can be worked out in two steps. In the first step $S_{\mathrm{vN}}$
is maximized for fixed arbitrary values of the variables $A_{i} $; in the
second step the outcome, already identified with $S_{\mathrm{th}}\left(
A_{i}\right)  $, is maximized with respect to the variables $A_{i}$ subject to
constraints imposed by the conservation laws and by the possible exchanges
between subsystems. The Second Law then appears merely as the \textit{second
step} of this maximization, which leads to the largest disorder.

Eqs. (\ref{009})-(\ref{011}) reduce to standard equations in thermodynamics.
In particular the expression (\ref{011}) of the multipliers $\gamma_{i}$
allows us to identify them with Callen's \textit{intensive variables} of
thermodynamics; for instance, the multiplier $\gamma$ associated with the
energy of a subsystem\ is the corresponding inverse temperature. Also, $\Psi$
is identified with a Massieu thermodynamic potential. Owing to the large
number of constituents of thermodynamic system, the statistical fluctuations
of the macroscopic variables are negligible, the probabilistic nature of
statistical physics which underlies thermodynamics becomes invisible, so that
thermodynamics appears as a deterministic theory. The thermodynamic entropy
$S_{\mathrm{th}}$, although it can be identified with a measure of our lack of
information about the system at the microscopic scale, as well as the
parameters $\gamma$ entering the density operator, reach an objective status
and become meaningful for a single macroscopic system.

\section{Elimination of the irrelevant variables}

In non-equilibrium statistical mechanics, a central problem consists in
predicting the values at the time $t$ of some set of variables $A_{i}$ from
the knowledge of their values at the initial time $t_{0}$. The set of relevant
observables $\hat{A}_{i}$ now contains not only the constants of the motion
and the quantities controlled or observed experimentally, but also other
macroscopic variables, chosen in such a way that the variables $A_{i}$ will
eventually be coupled only to one another within a good approximation. We
shall return below to the choice of this set, which for the time being we
leave unspecified.

A general solution of this inference problem is provided by the
\textit{projection method} of Nakajima and Zwanzig, which we only sketch here;
for a simple introduction, see Balian (1999). Since at the microscopic level
the dynamics is generated by the Liouville--von~Neumann equation (\ref{002}),
the equations of motion of $A_{i}\left(  t\right)  $ should be generated from
(\ref{002}) and (\ref{005}). We thus need first to transform the
\textit{initial conditions} on $A_{i}\left(  t_{0}\right)  $ into initial
conditions on $\hat{D}\left(  t_{0}\right)  $. Since at the initial time
$t_{0}$ nothing else than the set $A_{i}\left(  t_{0}\right)  $ is specified,
the least biased choice for $\hat{D}\left(  t_{0}\right)  $ is given by the
maximum entropy criterion, in the form (\ref{006}), (\ref{007}).

From this initial state $\hat{D}\left(  t_{0}\right)  =\hat{D}_{\mathrm{R}%
}\left(  t_{0}\right)  $, one derives the state $\hat{D}\left(  t\right)  $ at
an arbitrary time $t$ by solving the equation of motion (\ref{002}). The
relevant variables $A_{i}\left(  t\right)  $ are then found from $\hat
{D}\left(  t\right)  $ through (\ref{005}), which formally solves our
inference problem. We give below a more explicit expression for this\ answer.

The von~Neumann entropy (\ref{004}) associated with $\hat{D}\left(  t\right)
$ remains \textit{constant} in time, as readily checked from the unitarity of
the evolution (\ref{002}). This means that, if one considers \textit{all
possible observables}, \textit{no information is lost} during the evolution.
However, in general, $\hat{D}\left(  t\right)  $ does not keep the exponential
form (\ref{006}), which involves only the relevant observables $\hat{A}_{i}$.
Following the procedure of sections 4 and 5, we can evaluate the lack of
information associated with the knowledge of the variables $A_{i}\left(
t\right)  $ only. It is measured by the relevant entropy (\ref{010}), where
the multipliers $\gamma_{i}\left(  t\right)  $ are now time-dependent and are
expressed in terms of the set $A_{i}\left(  t\right)  $ through (\ref{009}) or
(\ref{011}). By construction, we have
\begin{align}
S_{\mathrm{R}}\left[  A_{i}\left(  t\right)  \right]   &  \geq S_{\mathrm{vN}%
}\left[  \hat{D}\left(  t\right)  \right]  =S_{\mathrm{vN}}\left[  \hat
{D}\left(  t_{0}\right)  \right] \tag{12}\label{012}\\
&  =S_{\mathrm{vN}}\left[  \hat{D}_{\mathrm{R}}\left(  t_{0}\right)  \right]
\equiv S_{\mathrm{R}}\left[  A_{i}\left(  t_{0}\right)  \right]  \text{
.}\nonumber
\end{align}
The fact that the relevant entropy is in general larger at the time $t$ than
at the initial time $t_{0}$ means that a part of our initial information on
the set $\hat{A}_{i}$ has leaked at the time $t$ towards irrelevant variables.
This \textit{loss of relevant information} characterizes a dissipation in the
evolution of the variables $A_{i}\left(  t\right)  $.

The construction of section 5 associates, at each time, a \textit{reduced
density operator} $\hat{D}_{\mathrm{R}}\left(  t\right)  $ with the set of
relevant variables $A_{i}\left(  t\right)  $. As regards these variables,
$\hat{D}_{\mathrm{R}}\left(  t\right)  $ is equivalent to $\hat{D}\left(
t\right)  $ :
\begin{equation}
\operatorname{Tr}\hat{D}_{\mathrm{R}}\left(  t\right)  \hat{A}_{i}%
=\operatorname{Tr}\hat{D}\left(  t\right)  \hat{A}_{i}=A_{i}\left(  t\right)
\text{\ ,}\tag{13}\label{013}%
\end{equation}
but it has the maximum entropy form (\ref{006}) and thus does not retain
information about the irrelevant variables. It is parameterized by the set of
multipliers $\gamma_{i}\left(  t\right)  $, in one-to-one correspondence with
the set of relevant variables $A_{i}\left(  t\right)  $. Regarding density
operators as points in a vector space, we can visualize (fig. 1) the
correspondence from $\hat{D}\left(  t\right)  $ to $\hat{D}_{\mathrm{R}%
}\left(  t\right)  $ as a \textit{projection} onto the manifold $\mathrm{R}$
of reduced states (\ref{006}). (The space of states $\hat{D}$\ can be endowed
with a natural metric (Balian et al, 1986) defined by $ds^{2}=-d^{2}%
S=\operatorname{Tr}\,d\hat{D}\,d\ln\hat{D}$; the correspondence\ $\hat
{D}\rightarrow\hat{D}_{\mathrm{R}}$ then appears as an orthogonal projection.)
We thus consider in parallel two time-dependent descriptions of the system,
the most detailed one by $\hat{D}\left(  t\right)  $ which accounts for all
observables, and the less detailed one by $\hat{D}_{\mathrm{R}}\left(
t\right)  $, or equivalently by the set $\gamma_{i}\left(  t\right)  $, which
accounts for the variables $A_{i}\left(  t\right)  $ only, and is sufficient
for our purpose.

\begin{figure}[h]
\centerline{\includegraphics*[width=9.5cm,keepaspectratio=true]{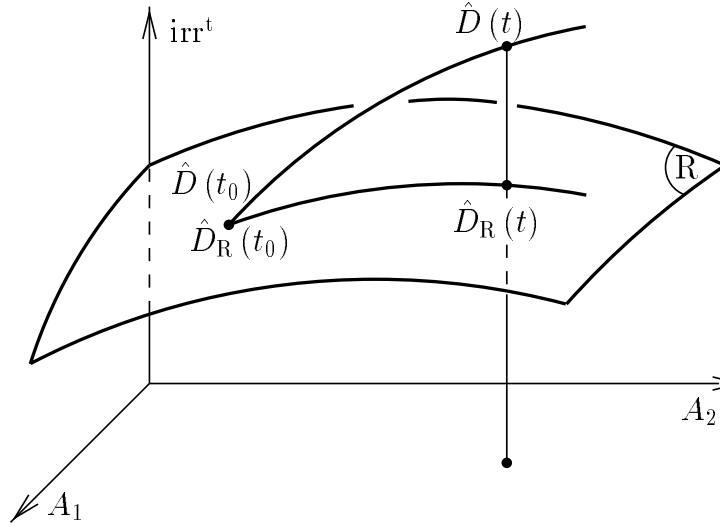}}\caption{\textit{The
reduction of the description.}{\small \ A state }$\hat{D}${\small \ of quantum
statistical physics is represented by a point in a vector space. Here the
first two axes schematize the relevant variables }$A_{i}=\operatorname{Tr}%
\hat{A}_{i}\hat{D}${\small , regarded as coordinates of }$\hat{D}${\small ,
while the third axis stands for its many irrelevant coordinates. The
trajectory of }$\hat{D}\left(  t\right)  ${\small \ is governed by the
Liouville--von~Neumann equation (\ref{002}). The reduced state }$\hat
{D}_{\mathrm{R}}\left(  t\right)  ${\small , equivalent to }$\hat{D}\left(
t\right)  ${\small \ as regards the relevant variables }$\left\{
A_{i}\right\}  ${\small \ but maximizing the von~Neumann entropy, is the
intersection of the plane }$A_{i}\left(  t\right)  =\operatorname{Tr}\hat
{A}_{i}\hat{D}${\small \ (represented here as a line) and the surface
}$\text{R}${\small \ parameterized by the set }$\left\{  \gamma_{i}\right\}
${\small \ according to (\ref{006}), (\ref{008}). The trajectory of }$\hat
{D}_{\mathrm{R}}\left(  t\right)  ${\small \ is obtained by projecting that of
}$\hat{D}\left(  t\right)  ${\small \ on }$\text{R}${\small . It starts from
the initial point }$\hat{D}_{\mathrm{R}}\left(  t_{0}\right)  =\hat{D}\left(
t_{0}\right)  ${\small \ and is governed by eq.(\ref{017}) or its
approximation (\ref{021}). The von~Neumann entropy is constant along the line
}$\hat{D}\left(  t\right)  ${\small , whereas for }$\hat{D}_{\mathrm{R}%
}\left(  t\right)  ${\small \ information is lost towards the irrelevant
degrees of freedom.}}%
\label{fig1}%
\end{figure}

Rather than following the motion of $\hat{D}\left(  t\right)  $ which involves
all the complicated irrelevant variables, we wish to eliminate these
variables, i.e., to follow the motion of $\hat{D}_{\mathrm{R}}\left(
t\right)  $ on the surface $\mathrm{R}$. We thus wish to project the
Liouville--von~Neumann trajectory of $\hat{D}\left(  t\right)  $ onto
$\mathrm{R}$. Once the operators $\hat{D}$ are regarded as elements of the
\textit{vector space of states}, it is natural to regard the operators
$\hat{A}$ as elements of the \textit{dual} vector space, the \textit{space of
observables}, the scalar product being defined by (\ref{001}) and noted as
\begin{equation}
\mbox{\boldmath{$($}}\hat{A}\mbox{\boldmath{$;$}}\hat{D}%
\mbox{\boldmath{$)$}}\equiv\operatorname{Tr}\hat{A}\hat{D}=\left\langle
\hat{A}\right\rangle _{\hat{D}}\text{ .}\tag{14}\label{014}%
\end{equation}
Linear transformations in either vector spaces $\left\{  \hat{D}\right\}  $ or
$\left\{  \hat{A}\right\}  $ are represented by \textquotedblleft
superoperators\textquotedblright\ acting either on their right or their left
side. In particular the equation of motion (\ref{002}) can be rewritten as
\begin{equation}
\frac{d\hat{D}}{dt}=\mathcal{L}\hat{D}\text{\ ,}\tag{15}\label{015}%
\end{equation}
in terms of the \textit{Liouvillian superoperator} which transforms $\hat{D}$
into $\mathcal{L}\hat{D}\equiv$\newline$\left[  \hat{H},\hat{D}\right]
/i\hbar$\ . The projection from $\hat{D}$ to $\hat{D}_{\mathrm{R}}%
=\mathcal{P}\hat{D}$ is implemented by means of the \textit{projection
superoperator}
\begin{equation}
\mathcal{P}=\hat{D}_{\mathrm{R}}\otimes\hat{I}+\frac{\partial\hat
{D}_{\mathrm{R}}}{\partial A_{i}}\otimes\left(  \hat{A}_{i}-A_{i}\hat
{I}\right)  \text{\ ,}\tag{16}\label{016}%
\end{equation}
where $\hat{D}_{\mathrm{R}}$, expressed by (\ref{006}), is regarded as a
function of the set $\gamma_{i}$ directly and through $\Psi$, and hence as a
function of the set $A_{i}$ through (\ref{011}). A superoperator which is a
tensor product $\hat{A}\otimes\hat{B}$ of operators acts on $\hat{C}$ as
$\hat{A}\operatorname*{Tr}\left(  \hat{B}\hat{C}\right)  $ on the right side
and as $\operatorname*{Tr}\left(  \hat{C}\hat{A}\right)  \hat{B}$ on the left
side. Applied on the right side, $\mathcal{P}$ leaves $\hat{D}_{\mathrm{R}}$
and its first-order derivatives $\partial\hat{D}_{\mathrm{R}}/\partial
\gamma_{i}$ or $\partial\hat{D}_{\mathrm{R}}/\partial A_{i}$ invariant.
Applied on the left side, it leaves the operators $\hat{A}_{i}$ and $\hat{I}$
invariant. We note the complementary projection superoperator on the
irrelevant subspace as $\mathcal{Q=J-P}$ where $\mathcal{J}$ is the unit
superoperator. We can then replace (\ref{015}), using (\ref{016}), by coupled
equations for the two projected states, $\hat{D}_{\mathrm{R}}=\mathcal{P}%
\hat{D}$, which depends on time through $\hat{D}$ and $\mathcal{P}$,\ and
$\mathcal{Q}\hat{D} $. The elimination of the irrelevant variables is then
achieved, explicitly though formally, by eliminating $\mathcal{Q}\hat{D}$.

We thus find an integro-differential equation of the form
\begin{equation}
\frac{d\hat{D}_{\mathrm{R}}}{dt}=\mathcal{PL}\hat{D}_{\mathrm{R}}+\int_{t_{0}%
}^{t}dt^{\prime}\mathcal{M}\left(  t,t^{\prime}\right)  \hat{D}_{\mathrm{R}%
}\left(  t^{\prime}\right)  \text{\ ,}\tag{17}\label{017}%
\end{equation}
which involves a \textit{memory kernel}
\begin{equation}
\mathcal{M}\left(  t,t^{\prime}\right)  =\mathcal{P}\left(  t\right)
\mathcal{L\ W}\left(  t,t^{\prime}\right)  \mathcal{LP}\left(  t^{\prime
}\right) \tag{18}\label{018}%
\end{equation}
acting in the time-dependent relevant space. This kernel depends on the
\textit{evolution superoperator} $\mathcal{W}$ \textit{in the irrelevant
space}, itself defined by
\begin{equation}
\left[  \frac{d}{dt}-\mathcal{Q}\left(  t\right)  \mathcal{L}\right]
\mathcal{W}\left(  t,t^{\prime}\right)  =\mathcal{Q}\left(  t\right)
\delta\left(  t-t^{\prime}\right)  \text{\ .}\tag{19}\label{019}%
\end{equation}
Eq. (\ref{017}) is equivalent to the equation of motion for the relevant
variables $A_{i}\left(  t\right)  =\mbox{\boldmath{$($}}\hat{A}_{i}%
\mbox{\boldmath{$;$}}\hat{D}\left(  t\right)
\mbox{\boldmath{$)$}}=\mbox{\boldmath{$($}}\hat{A}_{i}%
\mbox{\boldmath{$;$}}\hat{D}_{\mathrm{R}}\left(  t\right)
\mbox{\boldmath{$)$}}$, namely :
\begin{equation}
\frac{dA_{i}}{dt}=\mbox{\boldmath{$($}}\hat{A}_{i}%
\mbox{\boldmath{$;$}}\mathcal{PL}\hat{D}_{\mathrm{R}}%
\mbox{\boldmath{$)$}}+\int_{t_{0}}^{t}dt^{\prime}\mbox{\boldmath{$($}}\hat
{A}_{i}\mbox{\boldmath{$;$}}\mathcal{M}\left(  t,t^{\prime}\right)  \hat
{D}_{\mathrm{R}}\left(  t^{\prime}\right)  \mbox{\boldmath{$)$}}\text{\ ,}%
\tag{20}\label{020}%
\end{equation}
or for their conjugate set $\gamma_{i}\left(  t\right)  $. The first term
describes the direct coupling between relevant observables. The second one,
which results from their indirect coupling through the irrelevant ones, is a
\textit{retarded}, \textit{memory effect} depending on the past history of the
$A_{i}$'s.

We have already shown that, between the times $t_{0}$ and $t$, some
information about the relevant variables is transferred towards the irrelevant
ones, which keep memory of the initial conditions. This transfer of
information is measured by (\ref{012}). The \textit{dissipation}, that is, the
time-derivative of the relevant entropy, is obtained from (\ref{002}),
(\ref{006}), (\ref{017}) as
\begin{equation}
\frac{dS_{\mathrm{R}}}{dt}=\sum_{i}\gamma_{i}\left(  t\right)  \int_{t_{0}%
}^{t}dt^{\prime}\mbox{\boldmath{$($}}\hat{A}_{i}%
\mbox{\boldmath{$;$}}\mathcal{M}\left(  t,t^{\prime}\right)  \hat
{D}_{\mathrm{R}}\left(  t^{\prime}\right)  \mbox{\boldmath{$)$}}\text{\ .}%
\tag{21}\label{021}%
\end{equation}
It is a \textit{retardation effect}, associated with the history of the
coupling with the irrelevant observables. (The first term of (\ref{017}) does
not contribute.)

Till now, we have kept the relevant observables $\hat{A}_{i}$ arbitrary.
Although exact, the equation (\ref{020}) is in general unworkable due to
retardation. If we wish to use it in practice, we should make a suitable
choice of the set $\hat{A}_{i}$. Suppose this choice implies that the memory
time over which $\mathcal{M}\left(  t,t^{\prime}\right)  $ is sizeable is
short compared to the characteristic times of the set $A_{i}\left(  t\right)
$. This property occurs if (\ref{018}) involves very many irrelevant degrees
of freedom which evolve rapidly and interfere destructively. We can then
replace $\hat{D}_{\mathrm{R}}\left(  t^{\prime}\right)  $ by $\hat
{D}_{\mathrm{R}}\left(  t\right)  $ and $t_{0}$ by $-\infty$ in (\ref{017}).
In this approximation the evolution of the relevant variables is governed by a
mere differential equation at the time $t$%
\begin{equation}
\frac{d\hat{D}_{\mathrm{R}}}{dt}\simeq\mathcal{PL}\hat{D}_{\mathrm{R}%
}+\mathcal{K}\hat{D}_{\mathrm{R}}\text{\ ,}\tag{22}\label{022}%
\end{equation}
where the dissipative kernel $\mathcal{K}$ is defined by
\begin{equation}
\mathcal{K}\left(  t\right)  =\int_{-\infty}^{t}dt^{\prime}\mathcal{M}\left(
t,t^{\prime}\right)  \text{\ .}\tag{23}\label{023}%
\end{equation}

In this short-memory approximation, the dissipation (\ref{021}) reduces to
\begin{equation}
\frac{dS_{\mathrm{R}}}{dt}=\sum_{i}\gamma_{i}\mbox{\boldmath{$($}}\hat{A}%
_{i}\mbox{\boldmath{$;$}}\mathcal{K}D_{\mathrm{R}}%
\mbox{\boldmath{$)$}}\text{\ .}\tag{24}\label{024}%
\end{equation}
Although $S_{\mathrm{R}}\left(  t\right)  >S_{\mathrm{R}}\left(  t_{0}\right)
$, nothing prevents (\ref{021}) to be negative at some times for an arbitrary
choice of the set $\hat{A}_{i}$; at such times, relevant information (or
order) which had previously been transferred to the irrelevant set and
memorized there is recovered. Such a phenomenon indicates that the relevant
variables have not been properly selected since they are sensitive to memory
stored in the discarded variables.

However, if the short-memory approximation (\ref{022}) holds, the relevant
information is \textit{irretrievably lost}: the dissipation rate (\ref{024})
is positive. We understand this fact by noting that $\hat{D}_{\mathrm{R}%
}\left(  t\right)  $ depends only on $\hat{D}_{\mathrm{R}}\left(  t-\Delta
t\right)  $ if the time-scale $\Delta t$ characterizing the variations of
$\hat{D}_{\mathrm{R}}$ or $S_{\mathrm{R}}$ in (\ref{022}) or (\ref{024}) can
be treated as infinitesimal, but is large compared to the memory time of
$\mathcal{M}\left(  t,t^{\prime}\right)  $; the much earlier initial condition
$\hat{D}_{\mathrm{R}}\left(  t_{0}\right)  $ is forgotten. Taking then $t$ as
the initial time in eq. (\ref{012}), we find $S_{\mathrm{R}}\left(  t\right)
\geq S_{\mathrm{R}}\left(  t-\Delta t\right)  $ as anticipated. This
\textit{entropy production} measures a \textit{continuous flow of order} or of
information from the relevant to the irrelevant variables, which is generated
by the evolution of the system.

Altogether, if we choose the set of relevant observables $\hat{A}_{i}$ in such
a way that the memory time associated with the dynamics of the irrelevant
variables through $\mathcal{M}\left(  t,t^{\prime}\right)  $ is much shorter
than the time scales associated with the motion of the relevant variables
$A_{i}\left(  t\right)  $, the latter variables evolve \textit{autonomously}.
They obey the \textit{Markovian} equation (\ref{022}), and the associated
entropy (\ref{024}) \textit{increases}. The quasi-instantaneous kernel
$\mathcal{K}$ accounts for the dissipative effects induced on the variables
$A_{i}\left(  t\right)  $ by the eliminated ones. An adequate choice of
observables $\hat{A}_{i}$, for which equations of motion of the form
(\ref{022}) involving through $\mathcal{K}$ a negligible memory time are valid
within a given precision, is thus not left to us. It is \textit{imposed by the
microscopic dynamics} through the possibility of separating the time scales of
$A_{i}\left(  t\right)  $ and $\mathcal{M}\left(  t,t^{\prime}\right)  $. this
criterion does not depend on the observers, but only on the microscopic
Hamiltonian and the considered conditions. For finite systems, eq. (\ref{022})
still keeps a subjective aspect even if the short-memory approximation is
justified, because it deals with probabilities or equivalently with variables
$A_{i}\left(  t\right)  $ that are expectation values.\ However, for large
systems, when statistical fluctuations are negligible, the equations of motion
for the set $A_{i}\left(  t\right)  $ reach an objective status and can
describe an individual system, in the same way as the equilibrium properties
in thermodynamics (end of section 6). In such a situation, the increase of
relevant entropy measured by the dissipation rate (\ref{024}) also becomes
independent of the observers, although it is microscopically interpreted as a
loss of information. It quantifies the fact that the irrelevant variables act
only through the quasi-instantaneous kernel $\mathcal{K}$, and that nothing is
stored in them that could flow back towards the relevant set within any
reasonable delay.

Many semi-phenomenological approaches can be recovered from microphysics
through the above elimination of a set of irrelevant variables that produce no
memory effects on the relevant ones. We discuss two examples in sections 8 and 9.

\section{Thermodynamics of irreversible processes}

Section 6 was devoted to what is usually referred to as ``thermodynamics'',
but should rather be called ``thermo\textit{statics}''\ since its Laws apply
to the comparison of initial and the final state, not to the process itself
which leads from one to the other. Thermo\textit{dynamics} proper provides the
general laws that govern the time-dependence of the macroscopic variables, in
compound systems of the type described at the beginning of section 6. These
laws pertain to the thermodynamic regime, which must be sufficiently slow so
that the variables $A_{i}\left(  t\right)  $ that characterize the macroscopic
state at each time are the same \textit{conservative local variables} as in
section 6. However, the subsystems may now be volume elements, treated as
infinitesimal but sufficiently large so that each one remains nearly at
equilibrium at any time. Thus the variables $A_{i}$ include the energies of
the subsystems (or the energy density $\rho_{E}\left(  \mathbf{r}\right)  $ in
each volume element) for thermal processes; they include the numbers of
constitutive particles (or their local density $\rho_{N}\left(  \mathbf{r}%
\right)  $) for diffusive or chemical processes. They also include the density
of momentum $\rho_{\mathbf{P}}\left(  \mathbf{r}\right)  $ in hydrodynamics.

The equations of motion of thermodynamics couple the time-derivatives of the
conservative variables $A_{i}$ to their fluxes, which in turn are related to
the gradients of the intensive variables (defined as in the end of section 6).
They describe phenomena such as \textit{relaxation} or \textit{transport}, in
physics, mechanics or chemistry. The time-scales associated with this set of
macroscopic equations are large compared to the time-scales over which the
subsystems reach equilibrium. Indeed, owing to their conservative nature, the
variables $A_{i}$ can change only if a transfer occurs from one subsystem to
another, and the couplings which govern such transfers are weak compared to
the interactions within each subsystem which ensure local equilibrium.

Hence, in the thermodynamic regime, the relevant variables of microscopic
statistical physics should be identified with the above thermodynamic
variables $A_{i}$. In the projected equations (\ref{020}) that they obey,
there is a clear separation of time scales and the short-memory approximation
(\ref{022}), (\ref{023}) is valid. The resulting coupled instantaneous
equations for the set $A_{i}$ can then be identified with the phenomenological
equations of thermodynamics.

Since the relevant observables $\hat{A}_{i}$ are the same as in section 6,
their associated relevant entropy $S_{\mathrm{R}}\left(  A_{i}\right)  $ is
again identified with the thermodynamic entropy $S_{\mathrm{th}}\left(
A_{i}\right)  $; its partial derivatives are related to the local intensive
variables such as the local temperature, chemical potential or hydrodynamic
velocity. Its rate of change, given by statistical physics as (\ref{024}), is
positive. This property is identified with the macroscopic
\textit{Clausius--Duhem inequality} $dS_{\mathrm{th}}/dt\geq0$, which can
therefore be interpreted microscopically as the fact\ that some amount of
information, or some order attached to the conservative variables is
continuously lost during a dissipative thermodynamic process towards the
non-conservative variables.

\section{Boltzmann gas}

The first historic appearance of statistical mechanics was the kinetic theory
of dilute gases. In Boltzmann's approach, the state of the gas is represented
at each time by the density of particles $f\left(  \mathbf{r},\mathbf{p}%
,t\right)  $ in the $6$-dimensional single-particle phase space. Its evolution
is governed by the semi phenomenological \textit{Boltzmann equation}
\begin{equation}
\frac{\partial f}{\partial t}+\frac{\mathbf{p}}{m}\cdot\nabla_{\mathbf{r}%
}f=\mathcal{J}\left(  f\right)  \text{\ ,}\tag{25}\label{025}%
\end{equation}
the right side of which is the collision integral. It is local in space and time.

Boltzmann's description should be regarded as macroscopic. Indeed, in order to
describe at the microscopic scale a dilute gas, we should use the classical
limit of quantum statistical mechanics. We indicated in section 2 that the
probabilistic state is then represented by the density $D$ in the
$6N$-dimensional phase space, which encompasses not only $f\left(
\mathbf{r},\mathbf{p}\right)  $ but also the $2$-particle, ..., $N$-particle
correlation functions in the single-particle phase space. Their evolution is
governed by the equations of the BBGKY hierarchy, which are equivalent to the
Liouville equation (the classical limit of (\ref{002})) and which, in contrast
to Boltzmann's equation (\ref{025}), are invariant under time-reversal. Let us
identify $f\left(  \mathbf{r},\mathbf{p},t\right)  $ with our relevant set of
variables $A_{i}$, the index $i$ standing here for a point $\mathbf{r}$,
$\mathbf{p}$ in phase space. Following the procedure of section 7, we can
eliminate formally all the irrelevant variables, that is, all the correlations
between particles. The BBGKY hierarchy then gives rise to an
integro-differential equation of the type (\ref{020}) for $f\left(
\mathbf{r},\mathbf{p},t\right)  $. However, in a dilute gas, the memory time
entering the kernel $\mathcal{W}$, which describes the variations in time of
the correlations, is the \textit{duration of a collision}, since correlations
are created by two-particle collisions. This duration is much shorter than the
\textit{time between two successive collisions} of a particle, which governs
the evolution of $f\left(  \mathbf{r},\mathbf{p},t\right)  $. Likewise, the
size of the particles is much smaller than the mean free path. The
short-memory approximation is justified by this neat separation of scales. It
leads on the macroscopic time-scale to an instantaneous and local equation for
$f\left(  \mathbf{r},\mathbf{p},t\right)  $ of the type (\ref{022}), which is
identified with the Boltzmann equation (\ref{025}).

The relevant entropy $S_{\mathrm{B}}$ associated with the reduced Boltzmann
description, in terms of the single-particle density $f\left(  \mathbf{r}%
,\mathbf{p},t\right)  $, is readily found by noting that $D_{\mathrm{R}}$
factorizes as a product of identical contributions from each particle and that
the classical limit of a trace is an integral over phase space. We then get at
each time
\begin{equation}
S_{\mathrm{B}}\equiv S_{\mathrm{R}}\left(  f\right)  =h^{-3}\int
d^{3}\mathbf{r}\ d^{3}\mathbf{p}\ f\left(  \mathbf{r},\mathbf{p}\right)
\left[  1-\ln h^{3}\ f\left(  \mathbf{r},\mathbf{p}\right)  \right]  \text{
.}\tag{26}\label{026}%
\end{equation}
We recognize that this \textit{single-particle entropy} is directly related to
Boltzmann's $H$-functional, $H\equiv\int f\ \ln f$, within a sign and within
additive and multiplicative constants. Thus, \textit{Boltzmann's }%
$H$\textit{-theorem}, which expresses that $H$ does not increase when $f$
evolves according to Boltzmann's equation (\ref{025}), is now interpreted as a
continuous transfer of information from the single-particle data $f\left(
\mathbf{r},\mathbf{p}\right)  $ to correlations which are built up by each
collision. In this process, $S\left(  D\right)  $ remains constant, expressing
that no information would have been lost if we had been able to deal with the
full BBGKY hierarchy, including correlations between a very large number of particles.

Boltzmann's reduced description in terms of $f\left(  \mathbf{r}%
,\mathbf{p}\right)  $ is more detailed than the hydrodynamic or thermodynamic
description of the gas in terms of the densities $\rho_{N}\left(
\mathbf{r}\right)  $, $\rho_{\mathbf{P}}\left(  \mathbf{r}\right)  $,
$\rho_{E}\left(  \mathbf{r}\right)  $ of the conserved quantities (particle
number, momentum, energy). It depends at each time on one function $f$ of six
variables rather than on five functions $\rho$ of three variables, which are
obtained from $f\left(  \mathbf{r},\mathbf{p}\right)  $ by integration over
$\mathbf{p}$ with weights $1$, $\mathbf{p}$ and $p^{2}/2m$, respectively.
Boltzmann's description thus remains valid in circumstances where the
hydrodynamic description fails, such as shock waves and boundary layers where
space variations are so rapid that local equilibrium does not exist, or very
dilute gases in ballistic regimes where collisions are too rare to ensure thermalization.

Accordingly, \textit{Boltzmann's entropy} $S_{\mathrm{B}}$ defined by
(\ref{026}) should not be confused with the \textit{thermodynamic entropy}
$S_{\mathrm{th}}$, which for a gas is a function of the $\rho$'s given by the
Sackur--Tetrode formula. Nor should the $H$\textit{-theorem} for
$S_{\mathrm{B}}$ be confused with the \textit{Second Law} or with the
\textit{Clausius--Duhem inequality} for $S_{\mathrm{th}}$. The latter is valid
in the local equilibrium regime only (but for any material, in contrast to the
$H$-theorem). For dilute gases in the ballistic regime, only $S_{\mathrm{B}}$
is relevant, $S_{\mathrm{th}}$ is useless.

Since Boltzmann's description is less reduced than the hydrodynamic one,
$S_{\mathrm{B}}$ is in general smaller than $S_{\mathrm{th}}$. Actually,
maximization of (\ref{026}) under local constraints on $\rho_{N}$,
$\rho_{\mathbf{P}}$ and $\rho_{E}$ provides for $f\left(  \mathbf{r}%
,\mathbf{p}\right)  $ a Maxwellian form, i.e., $\ln f\left(  \mathbf{r}%
,\mathbf{p}\right)  $ is a linear function of $\mathbf{p}$ and $p^{2}/2m$
describing local equilibrium. When a gas is enclosed in a vessel, its
irreversible evolution, which is governed by (\ref{025}), leads it in a first
stage to a state close to local \ equilibrium; Boltzmann's entropy
$S_{\mathrm{B}}$ increases and eventually reaches $S_{\mathrm{th}}$.
Information is thus lost, not only towards the \textit{correlations} which are
not observed, but also through the evolution of $f\left(  \mathbf{r}%
,\mathbf{p}\right)  $ towards a maximum entropy \textit{Maxwellian} form. In
the second stage, after the local equilibrium regime is reached, the collision
term in (\ref{025}) is dominated by the small deviation of $f\left(
\mathbf{r},\mathbf{p}\right)  $ from this form. The Chapman--Enskog method,
based on the elimination of this deviation (which is formally similar to the
elimination of $\mathcal{Q}\hat{D}_{\mathrm{R}}$ that leads to (\ref{017})
then to (\ref{022})), then provides the thermodynamic equations for the gas
(Fourier and Navier--Stokes equations) as consequences of the Boltzmann
equation. In this regime the two descriptions become equivalent;
$S_{\mathrm{B}}$ and $S_{\mathrm{th}}$, which remain nearly equal, both
increase towards the global equilibrium entropy. The information which is
being lost during this second process is the one associated with
\textit{non-uniformity} in space of the densities $\rho$.

\section{Irreversibility and the multiplicity of entro\-pies}

The discussion of section 7 and the two examples of sections 8 and 9 elucidate
the \textit{paradox of irreversibility}, stated at the end of the XIXth
century by Loschmidt, Zermelo and Poincar\'{e}: How can the reversible
Liouville--von~Neumann equations of motion, which underlie the behaviour of
any system, give rise to the irreversible evolutions that we observe at our
scale? The answer is based on information theory. Macroscopic objects are so
complex at the microscopic scale that we can control or observe only an
extremely tiny proportion of the entire set of variables. These experimental
variables are part of a set $A_{i}$, the set of relevant variables, that we
may theoretically handle (the thermodynamic densities $\rho$ in section 8, the
single-particle density $f$ in section 9). Irreversibility is then associated
with an \textit{irretrievable leak of information} (\textit{or of order}) from
the set $A_{i}$ towards the irrelevant variables, extremely numerous and
completely out of reach. The time-derivative of the relevant entropy
$S_{\mathrm{R}}\left(  A_{i}\right)  $ measures this rate of dissipation.

The fact that order which is lost for us within the irrelevant degrees of
freedom never returns to the relevant set is a consequence of the occurrence
of \textit{two time scales in the microscopic dynamics}. The short memory-time
of the kernel $\mathcal{M}$ in (\ref{020}) is associated with the very large
number of irrelevant variables and with the destructive interference of their
evolutions. The longer time-scale which characterizes the motion of the
relevant set $A_{i}$, including the experimentally accessible variables, is
associated with some specificities of the dynamics: local conservation laws
for the thermodynamic regime, large ratio between the interparticle distance
and the range of forces for gases in the ballistic regime. This separation of
time scales provides, within a good approximation, instantaneous equations
(\ref{022}) for the relevant set which, in contrast to the microscopic
equations, are non-linear and irreversible. It should be stressed that the
short memory in the exact equation (\ref{020}) is sufficient to ensure
dissipation, and that the huge number of irrelevant variables also implies
that the Poincar\'{e} recurrence time is large beyond any imagination.
Moreover, we have seen that, although the underlying concepts of probability,
expectation value and information have a subjective nature, the resulting
properties for the variables $A_{i}\left(  t\right)  $ and for the relevant
entropy $S_{\mathrm{R}}\left(  t\right)  $ become objective, depending only on
the system in hand when this system is macroscopic.

Irreversibility, when analyzed at the microscopic scale, is a
\textit{statistical} phenomenon. We have seen that probabilities underlie the
very idea of information, a loss of which is identified with dissipation. An
individual trajectory in classical mechanics, a time-dependent pure state in
quantum mechanics, cannot display irreversibility, since they encompass all
the degrees of freedom. Boltzmann already recognized this fact in kinetic
theory, when he stressed the idea that, even when one considers a single
trajectory in the $6N$-dimensional phase space, its initial point should be
regarded as a \textit{typical} configuration belonging to some
\textit{ensemble} of macroscopically similar configurations. The logarithm of
the number of these configurations is the initial entropy. The final point of
the trajectory is likewise a typical configuration of a much larger set, all
elements of which describe equilibrium configurations. The larger size of this
set provides a quantitative interpretation of the entropy increase.

Regarding the relevant entropy as missing information has led us to assign to
a given process several entropies, associated with different levels of
description. For a gas we have thus characterized in section 9 the coarseness
of the most detailed description, of the Boltzmann description and of the
thermodynamic description by the von~Neumann entropy, the Boltzmann entropy
and the thermodynamic entropy, respectively. In order to understand better the
mechanism which gives rise to the irreversibility of the dynamics of a
classical gas, we can introduce a set of \textit{nested reduced descriptions}
(Mayer and Mayer, 1977). We start from the complete description in terms of
the density $D$ in the $6N$-dimensional phase space, which obeys the Liouville
equation. The $n$-th reduced description is defined by dropping all
information contained in the correlations of more than $n$ particles.
Boltzmann's description is the first reduced description; the second one
accounts for the two-particle correlations created by the collisions, but not
for the higher order correlations; and so on. At each level, there are reduced
equations of motion resulting from the BBGKY set. The associated
\textit{hierarchy of relevant entropies} $S_{1}\equiv S_{\mathrm{B}}$, $S_{2}%
$, \ldots, $S_{n}$, \ldots, $S_{N}\equiv S\left(  D\right)  $ satisfy the
inequalities
\begin{equation}
S_{\mathrm{eq}}\geq S_{\mathrm{th}}\geq S_{1}\geq S_{2}\geq\ldots\geq
S_{n}\geq\ldots\geq S\left(  D\right)  \text{ ,}\tag{27}\label{027}%
\end{equation}
which express an increasing content of information. Their time-dependence,
qualitatively shown in fig. 2, reflects the \textit{cascade} through which
correlations of more and more particles are created by successive collisions.
Since two particles that have already undergone a collision have little chance
to collide again for large $N$, due to the fact that the mean free path is
much longer than the interparticle distance, information flows out from
$f\left(  \mathbf{r},\mathbf{p}\right)  $ first to $2$-particle correlations,
then from $2$-particle to $3$-particle correlations, and so on.

\begin{figure}[h]
\centerline{\includegraphics*[width=9.5cm,keepaspectratio=true]{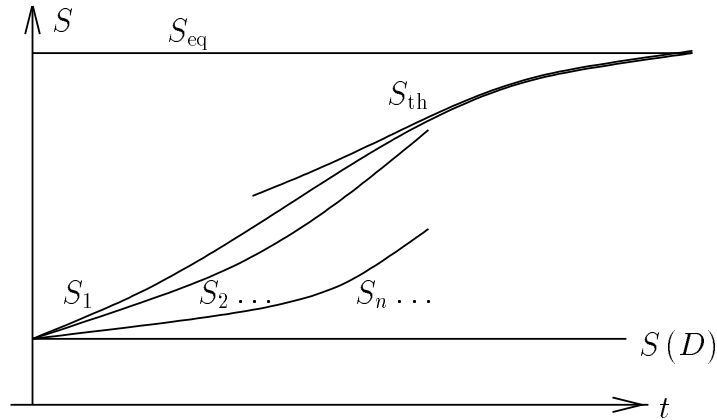}}\caption{\textit{Hierarchy
of entropies in a gas.}{\small \ Larger and larger entropies are associated
with less and less detailed descriptions. The entropy }$S\left(  D\right)
${\small \ of the complete description remains constant for an isolated
system. If the initial state is uncorrelated, Boltzmann's entropy }$S_{1}%
${\small \ increases (}$H${\small -theorem) from }$S\left(  D\right)
${\small \ to the equilibrium thermostatic entropy }$S_{\mathrm{eq}}${\small ,
evaluated for the initial values of the energy, particle number and density of
the gas. It is smaller than the thermodynamic entropy }$S_{\mathrm{th}}%
${\small , a function of the local densities of energy, particles and
momentum, but both remain nearly equal after the distribution }$f\left(
\mathbf{r},\mathbf{p},t\right)  ${\small \ has become Maxwellian (local
equilibrium regime). The entropies }$S_{2},S_{3}\cdots S_{n}$%
{\small \ accounting for }$2${\small -, }$3${\small -,}$\cdots n$%
{\small -particle correlations evolve like }$S_{1}${\small ; they are at each
time smaller and smaller with }$n${\small \ and increase later and later.}}%
\label{fig2}%
\end{figure}

Figure 2 also exhibits a non-uniform convergence. If $N$ is fixed, $S_{n}$
tends to the constant $S\left(  D\right)  $ at any time when $n$ increases so
as to reach $N$. However, under physical conditions, we have always $N\gg n$,
even if we let $n\gg1$. In this case, $S_{n}$ tends for any fixed $n$ and for
long times to the equilibrium entropy $S_{\mathrm{eq}}$. Thus, the amount of
order $S_{\mathrm{eq}}-S\left(  D\right)  $ which is in excess in the initial
non-equilibrium state is eventually dissipated within the \textit{correlations
of an infinite number of particles}. This property is consistent with the way
$D\left(  t\right)  $ converges for large $t$ towards the equilibrium state
$D_{\mathrm{eq}}$ of a perfect gas. Actually, for finite $N$, $D\left(
t\right)  $ cannot reach $D_{\mathrm{eq}}$ since their entropies differ.
However, if $N$ is made large first, $D\left(  t\right)  $ becomes equivalent
for large $t$ to $D_{\mathrm{eq}}$ as regards $f\left(  \mathbf{r}%
,\mathbf{p}\right)  $ (a Maxwellian distribution) and as regards all the
correlations of a \textit{finite} number $n$ of particles (these correlations
are negligible).

Note that the entropies of the above hierarchy are purely theoretical objects.
The only ones that experiments may reach, more or less indirectly, are
$S_{\mathrm{B}}$, $S_{\mathrm{th}}$ and $S_{\mathrm{eq}}$, depending on the circumstances.

\section{Spin echoes}

Let us return to the paradox of irreversibility. Consider a gas, enclosed in a
vessel and prepared at the initial time $t_{0}=0$ in an off-equilibrium state
(for instance with a non-uniform density). At the time $\tau$, it has reached
equilibrium. If we were able to reverse suddenly the velocities of all the
particles, we would recover the initial state at the time $2\tau$. Such a
violation of the arrow of time in thermodynamics cannot be experimentally
achieved, because the needed inversion of the individual velocities is not
feasible. Anyhow, since the number of microscopic configurations associated
with the initial non-equilibrium state is enormously smaller than the number
of those associated with the final equilibrium state, the above hypothetical
reversal of velocities should be performed rigorously, which is unthinkable; a
small deviation would lead at the final time $2\tau$ to a configuration having
equilibrium features at the macroscopic scale, in contrast to the initial
off-equilibrium configuration.

Nevertheless, similar paradoxical evolutions which violate at a macroscopic
scale the laws of thermodynamics have been built up, through subtle
experiments of nuclear magnetism (Abragam and Goldman, 1982). The nuclear
spins $\mathbf{\hat{s}}_{i}$ ($i=1$, $2$, \ldots, $N$), the value $s$ of which
is a half-integer,\ lie in an external magnetic field permanently applied
along the $z$-axis. They undergo a Larmor precession around this axis with
angular frequency $\mbox{\boldmath{$\omega$}}_{0}$ along $z$ and proportional
to the field. The only quantities that can be observed in practice are the
components of the total magnetic moment $\mathbf{M}\left(  t\right)  $, equal
in dimensionless units to the expectation value over the density operator
$\hat{D}\left(  t\right)  $ of the total spin
\begin{equation}
\mathbf{\hat{M}}\equiv\sum_{i}\mathbf{\hat{s}}_{i}\text{ .}\tag{28}\label{028}%
\end{equation}
One can act on the system only by means of time-dependent external fields
coupled to $\mathbf{\hat{M}}$. One thus handles simultaneously all the spins,
which globally rotate. In particular, application at time $t$ of a brief
magnetic pulse along some axis during a controlled delay allows one to
suddenly perform any overall rotation, with a given angle $\Omega$ around a
given direction $\mathbf{\Omega}/\Omega$. This pulse modifies the density
operator $\hat{D}\left(  t\right)  $ into $\hat{U}D\left(  t\right)  \hat
{U}^{\dagger}$, where $\hat{U}=\exp\left(  i\mathbf{\Omega}\cdot
\mathbf{\hat{M}}\right)  $ is the unitary transformation describing the
considered rotation.

The Hamiltonian has the following form
\begin{equation}
\hat{H}=\hbar\omega_{0}\ \hat{M}_{z}+\sum_{i}\hbar\delta
\mbox{\boldmath{$\omega$}}_{i}\cdot\mathbf{\hat{s}}_{i}+\hat{V}_{\mathrm{ss}%
}+\hat{V}_{\mathrm{sl}}\text{ .}\tag{29}\label{029}%
\end{equation}
Its first term generates the \textit{Larmor precession} $d\mathbf{M}%
/dt=\mbox{\boldmath{$\omega$}}_{0}\times\mathbf{M}$ around $z$. The next terms
are responsible for three different mechanisms of relaxation of $\mathbf{M}$.

(i) The applied field is \textit{not quite uniform}. Each spin $\left\langle
\mathbf{\hat{s}}_{i}\right\rangle $, depending on its location, undergoes a
Larmor precession with angular frequency $\mbox{\boldmath{$\omega$}}_{0}%
+\delta\mbox{\boldmath{$\omega$}}_{i}$. If all the spins $\left\langle
\mathbf{\hat{s}}_{i}\right\rangle $ lie initially along the $x$-axis, they
precess at different speeds, nearly in the $xy$-plane, and thus get graduately
out of phase. Hence their sum $\mathbf{M}\left(  t\right)  $ spirals down to
zero in the vicinity of the $xy$-plane. If we denote as $\delta$ the
statistical fluctuation of $\delta\omega_{iz}$ (the average over $i$ of which
vanishes) the relaxation time for this process is $1/\delta$.

(ii) The \textit{spin-spin} contribution $\hat{V}_{\mathrm{ss}}$ to
(\ref{029}) is the sum of the two-body dipolar interactions. Under the
experimental conditions where $\omega_{0}$ is sufficiently large, the part of
$\hat{V}_{\mathrm{ss}}$\ which does not commute with $\hat{M}_{z}$ is
ineffective. These pairwise interactions play for the spins the same role as
the interparticle interactions for a gas. A ``flip-flop''\ process, which
changes the configuration $\uparrow\downarrow$ of a pair into $\downarrow
\uparrow$, is the equivalent of a collision, which changes the momenta of two
particles in a gas. This mechanism gradually creates correlations between two,
then three spins, and so on. By a process similar to the one described at the
end of section 10, the state of the $N$-spin system thus tends to a maximum
entropy Boltzmann--Gibbs distribution $\hat{D}_{\mathrm{eq}}\propto\exp\left(
-\beta\hat{H}\right)  $, apart from invisible many-spin correlations. The spin
temperature $1/\beta$ depends only on the initial energy of the spins. The
relaxation time for this process is traditionally denoted as $T_{2}$.

(iii) The \textit{spin-lattice} contribution $\hat{V}_{\mathrm{sl}}$ to
(\ref{029}) couples the nuclear spins to the other degrees of freedom (the
``lattice''), such as the phonons in a solid. It tends to thermalize the
nuclear spins, imposing them the lattice temperature. Its associated
relaxation time $T_{1}$ is much longer than the duration of the spin echo
experiments, owing to the weakness of the coupling between the nuclear spins
and the other variables of the material. We shall thus disregard this term.

The oldest type of spin echoes was discovered by Hahn soon after the birth of
nuclear magnetic resonance. Such spin echo experiments are performed on
liquids, the disorder of which smoothes out the interactions $\hat
{V}_{\mathrm{ss}}$. The Hamiltonian thus reduces to the first two Zeeman terms
of (\ref{029}). One starts from an equilibrium state, at a sufficiently low
temperature so that all spins are practically oriented along $z$. By a pulse
$\pi/2$ applied along $y$ just before the time $t_{0}=0$, one generates an
off-equilibrium pure \textit{initial state} $\hat{D}\left(  0\right)  $ with
magnetization $M_{x}\left(  0\right)  $ having the largest possible value
$Ns$, and $M_{y}=M_{z}=0$. As indicated above under (i), the magnetization
$\mathbf{M}\left(  t\right)  $ \textit{relaxes} under the effect of the
heterogeneity of the field. After a number of turns larger than $\omega
_{0}/\delta$, at some time $\tau$ larger than $1/\delta$ (but smaller than
$T_{2}$ and $T_{1}$), all three components of $\mathbf{M}$ have reached $0$.
If we imagine we have followed the motion at the microscopic scale, the state
$\hat{D}\left(  t\right)  $ has remained pure: each one of the $N$ vectors
$\left\langle \mathbf{\hat{s}}_{i}\right\rangle $ has kept the maximum length
$s$; its component along $z$ is negligible as $\delta/\omega_{0}$, while it
points out in some direction in the $xy$-plane. This direction is determined
by the local magnetic field, i.e., by $\delta\omega_{iz}$. Nothing seems to
distinguish this state from the reduced state $\hat{D}_{\mathrm{R}}$ which
describes thermodynamic equilibrium at large temperature. While the entropy
$S\left(  \hat{D}\right)  $ has kept its initial value $0$, the relevant
entropy $S\left(  \hat{D}_{\mathrm{R}}\right)  =S_{\mathrm{R}}\left(
\mathbf{M}=0\right)  $\ associated with $\mathbf{M}$ has reached the largest
possible value $N\ln\left(  2s+1\right)  $. The system seems \textquotedblleft
dead\textquotedblright\ since only $\mathbf{M}=0$ is observed in practice.

However, superimposed to the permanent magnetic field, a pulse $\pi$ along $x$
is applied at the time $\tau$. The components $\left\langle \hat{s}%
_{ix}\right\rangle $, $\left\langle \hat{s}_{iy}\right\rangle $, and
$\left\langle \hat{s}_{iz}\right\rangle \simeq0$ are quasi suddenly changed by
the corresponding unitary transformation into $\left\langle \hat{s}%
_{ix}\right\rangle $, $-\left\langle \hat{s}_{iy}\right\rangle $, and
$-\left\langle \hat{s}_{iz}\right\rangle \simeq0$. Hence, the spins which
precess faster than the average Larmor flow because $\delta\omega_{iz}>0$, and
which were therefore ahead at the time $\tau$ by some angle, are now behind at
the time $\tau+0$ by just the same angle. Likewise, the spins that precess
slower have been brought forward by the pulse. Hence, after a second time
lapse $\tau$, during which the precession goes on with the same local
frequency, all the spins reach the initial direction along $x$, and the total
magnetization $\mathbf{M}$ \textit{returns to its largest possible value} $Ns$
at the time $2\tau$. Between the times $\tau$ and $2\tau$, the initial order
has progressively been recovered, the relevant entropy $S_{\mathrm{R}}\left(
\mathbf{M}\right)  $ has decreased from $N\ln\left(  2s+1\right)  $ to $0$. A
single \textit{macroscopic manipulation}, the application of the pulse $\pi$
along $x$ at $\tau$, has been sufficient to produce an evolution which
\textit{violates the thermodynamic surmise}, as would have done the unfeasible
reversal of velocities for a gas. This paradoxical evolution can be understood
only by keeping track of \textit{all} the microscopic spin degrees of freedom.

In this spin echo experiment, the order about $\mathbf{M}$, lost between the
time $0$ and $\tau$ and recovered between $\tau$ and $2\tau$, is associated
with the \textit{correlation for each spin}, which during the motion is
established\ between the direction of the vector $\left\langle \mathbf{\hat
{s}}_{i}\right\rangle $ in the $xy$-plane and the value $\delta\omega_{iz}$ of
the $z$-component of the local magnetic field. The very large number
$\omega_{0}\tau/\hbar$ of revolutions of the spins that take place during the
time $\tau$ produces a configuration of directions of spins $\left\langle
\mathbf{\hat{s}}_{i}\right\rangle $ which seems to be just a sample picked up
at random within the truly dead ensemble $\hat{D}_{\mathrm{R}}$ describing
equilibrium. Even if we are given the full available information, to wit, the
seemingly dead density operator $\hat{D}\left(  \tau\right)  $ issued from
$\hat{D}\left(  0\right)  $ and the Hamiltonian $\hat{H}$ that generated the
evolution, we would have an extreme difficulty to uncover, among this huge
amount of data, the crucial correlations between $\left\langle \mathbf{\hat
{s}}_{i}\right\rangle $ and $\delta\omega_{iz}$. Our situation is the same as
that of the audience of a conjuring show, who are unable to detect how the
shuffling process has retained some hidden order in a pack of cards. The
experimentalist has also the same pieces of information, $\hat{D}\left(
\tau\right)  $ and $\hat{H}$, but like the conjurer he acknowledges therein
that the directions of the $\left\langle \mathbf{\hat{s}}_{i}\right\rangle $'s
are not random but are correlated with the $\delta\omega_{iz}$'s. Indeed, he
is aware of the initial conditions and of the history of the system which
destroyed the visible order, an information which is equivalent to the
knowledge of the correlations between $D\left(  \tau\right)  $ and $\hat{H}$,
but easier to analyse. Relying on this knowledge, he is able to bring back the
order buried in these correlations into the simple macroscopic degrees of
freedom $\mathbf{M}$.

He succeeds in this task owing to a specificity of the motion: it is
two-dimensional except during a pulse, and we can see that the rotation
$\hat{U}$ around the $x$-axis has an effect akin to a time reversal. Indeed,
noting that the final state is invariant under the rotation $\hat{U}=\hat
{U}^{\dagger}$ around $x$, we can write the evolution operator between the
times $0$ and $2\tau$ as
\begin{equation}
\hat{U}^{\dagger}\exp\left(  -i\hat{H}\tau/\hbar\right)  \hat{U}\exp\left(
-i\hat{H}\tau/\hbar\right)  =\exp\left(  -i\hat{U}^{\dagger}\hat{H}\hat{U}%
\tau/\hbar\right)  \exp\left(  -i\hat{H}\tau/\hbar\right)  \text{ .}%
\tag{30}\label{030}%
\end{equation}
Since the components $\omega_{ix}$ of the field have little effect, we can
write
\begin{equation}
\hat{U}^{\dagger}\hat{H}\hat{U}\simeq-\hat{H}\text{ .}\tag{31}\label{031}%
\end{equation}
This change of sign has the same effect as a \textit{time-reversal} taking
place during the second part of the process, which therefore brings us back to
the initial states at the time $2\tau$.

After the time $2\tau$, the evolution is exactly the same as after the initial
time $0$ and $\mathbf{M}\left(  t\right)  $ returns to zero. The operation can
then be \textit{iterated}, and the measurement of the length of $\mathbf{M}%
\left(  t\right)  $ provides a sequence of peaks. Their height slowly
decreases due to the other relaxation mechanisms that we have neglected on the
time scale $1/\delta$. Spin echoes have thus become a current technique in
nuclear magnetic resonance to determine relaxation times with precision and
thus explore matter.

At the epoch of their discovery, these spin echoes were regarded as somewhat
miraculous since they violated thermodynamics: They appeared as an exceptional
macroscopic phenomenon that can be explained only through the microscopic
dynamics. Full knowledge of $\hat{D}\left(  \tau\right)  $ and of $\hat{H}$ is
needed here, whereas the thermodynamic phenomena can be described in terms of
the reduced state $\hat{D}_{\mathrm{th}}$ and of $\hat{H}$ only. We may,
however, argue that the order which is retrieved was not very deeply hidden.
It lay at the time $\tau$ in the rather trivial set of correlations between
the orientation of each spin and the size of the corresponding local field. It
is natural to think that the relaxation induced by the \textit{spin-spin
interactions} should in actual fact be incurable, in the same way as the
collision-induced relaxation in a gas. Nevertheless, quite surprisingly,
spin-echo experiments of a different type have been performed, in which the
\textquotedblleft\textit{death}\textquotedblright\ of spins ($\mathbf{M}%
\rightarrow0$) caused after the relaxation time $T_{2}$ by the term $\hat
{V}_{\mathrm{ss}}$ of (\ref{029}) is followed by their \textquotedblleft%
\textit{resurrection}\textquotedblright\ (Abragam and Goldman, 1982). We only
sketch here the basic ideas. The considered experiments are performed on
solids, for which $\hat{V}_{\mathrm{ss}}$ is significant. The applied field is
sufficiently uniform so that the duration of the experiment is much shorter
than $1/\delta$. The Hamiltonian (\ref{029}) can therefore be simplified into
$\hat{H}=\hbar\omega_{0}\hat{M}_{z}+\hat{V}_{\mathrm{ss}}$. The experiment
begins as in Hahn's spin echoes by the preparation of an initial state with
$M_{x}=Ns$, $M_{y}=M_{z}=0$, then during the Larmor precession by its decay,
taking now place under the effect of $\hat{V}_{\mathrm{ss}}$. At a time $\tau$
larger than $T_{2}$, the magnetization is lost as above, but now the
relaxation has taken place (as in the case of gases) through a cascade of
flip-flop transitions involving more and more spins. The initial order has
dissolved into complicated many-spin correlations, and the density operator
\begin{equation}
\hat{D}\left(  \tau\right)  =e^{-i\hat{H}\tau/\hbar}\hat{D}\left(  0\right)
e^{i\hat{H}\tau/\hbar}\tag{32}\label{032}%
\end{equation}
is equivalent, as regards observables involving a finite number of spins, to a
\textquotedblleft dead\textquotedblright\ equilibrium state with large temperature.

After the time $\tau$, a sequence of suitably chosen brief pulses and lasting
time-dependent fields referred to as a ``\textit{magic sandwich}'', is applied
for a duration $\tau^{\prime}$. Thus, during this period, the Hamiltonian
becomes $\hat{H}+\hbar\mbox{\boldmath{$\omega$}}\left(  t\right)
\cdot\mathbf{\hat{M}}$ where $\mbox{\boldmath{$\omega$}}\left(  t\right)  $ is
proportional to the time-dependent applied field. The remarkable fact is the
possibility of choosing $\mbox{\boldmath{$\omega$}}\left(  t\right)  $ in such
a way that the unitary operator which describes the evolution between the
times $\tau$ and $\tau+\tau^{\prime}$ has approximately the form
\begin{equation}
T\exp\left\{  -i\int_{\tau}^{\tau+\tau^{\prime}}dt\left[  \hat{H}%
/\hbar+\mbox{\boldmath{$\omega$}}\left(  t\right)  \cdot\mathbf{\hat{M}%
}\right]  \right\}  \simeq\exp\left[  i\lambda\hat{H}\tau^{\prime}%
/\hbar\right] \tag{33}\label{033}%
\end{equation}
($T$ denotes time-ordering). The coefficient $\lambda$ is positive, so that,
as in eqs. (\ref{030}) (\ref{031}), the effect of the magic sandwich
$\mbox{\boldmath{$\omega$}}\left(  t\right)  $ is to replace $\tau^{\prime}$
by $-\lambda\tau^{\prime}$, thus \textit{changing effectively the arrow of
time} in the considered interval. The actual realization of eq. (\ref{033})
has needed great ingenuity, because the dipolar interactions contained in
$\hat{H}$\ involve pairwise interactions whereas the external operations are
represented by a mere field coupled to $\mathbf{\hat{M}}$\ only.

As an example, in one of the experiments performed,
$\mbox{\boldmath{$\omega$}}\left(  t\right)  $\ in (\ref{033}) describes a
magic sandwich consisting of three successive operations: (i) at the time
$\tau$, a pulse $\pi/2$\ along $y$ is exerted, represented by the operator
$U$; (ii) between the times $\tau$\ and $\tau+\tau^{\prime}$, a radiofrequency
field along $x$, with a frequency $\omega_{0}$\ and an intensity $\hbar
\omega^{\prime}$\ is applied; in the Larmor rotating frame, it provides a
contribution $\hbar\omega^{\prime}\hat{M}_{x}$\ to the Hamiltonian; (iii) at
the time $\tau+\tau^{\prime}$\ a pulse $-\pi/2$\ is applied along $y$. The
evolution operator (\ref{033}), written in the rotating frame, is thus
\begin{equation}
\hat{U}^{\dagger}\exp\left[  -i\left(  \hat{V}_{\mathrm{ss}}/\hbar
+\omega^{\prime}\hat{M}_{x}\right)  \tau^{\prime}\right]  \hat{U}=\exp\left(
-i\hat{U}^{\dagger}\hat{V}_{\mathrm{ss}}\hat{U}\tau^{\prime}/\hbar
-i\omega^{\prime}\hat{M}_{z}\tau^{\prime}\right)  \text{ .}\tag{34}\label{034}%
\end{equation}
If $\omega^{\prime}$\ is sufficiently large, the part of $\hat{U}^{\dagger
}\hat{V}_{\mathrm{ss}}\hat{U}$\ which does not commute with $M_{z}$\ is
negligible, and one finds
\begin{equation}
\hat{U}^{\dagger}\hat{V}_{\mathrm{ss}}\hat{U}\simeq-\frac{1}{2}\hat
{V}_{\mathrm{ss}}\text{ ,}\tag{35}\label{035}%
\end{equation}
which replaces here (\ref{031}). The last term in (\ref{034}) amounts to a
trivial Larmor rotation. Thus (\ref{033}) is satisfied (in the rotating frame)
with $\lambda=\frac{1}{2}$.

After the time $\tau+\tau^{\prime}$\ the evolution is again governed by the
sole Hamiltonian $\hat{H}$. Hence, provided $\tau^{\prime}$\ is larger than
$\tau/\lambda$, we can introduce a delay $\tau^{\prime\prime}$\ such that
$\tau-\lambda\tau^{\prime}+\tau^{\prime\prime}=0$, for which the full
evolution operator from $t=0$ to $t=\tau+\tau^{\prime}+\tau^{\prime\prime}$,
found from (\ref{032}) and (\ref{033}) is nearly the unit operator. The
magnetization $\mathbf{M}\left(  \tau+\tau^{\prime}+\tau^{\prime\prime
}\right)  $\ thus returns to its initial value $\mathbf{M}\left(  0\right)
$\ although it had previously decayed to zero during the time $\tau$. Here
again the arrow of time is challenged.

The features of this process are the same as in Hahn's echoes, in particular
the \textit{recovery of hidden order} and the need for explaining this
phenomenon to rely on \textit{all the microscopic degrees of freedom},
although the observations and the manipulations concern only the macroscopic
quantities $\mathbf{M}$. However, here, the order that is experimentally
retrieved had been transferred by the evolution towards extremely complicated
correlations of large numbers of spins, and the initial decay seemed genuinely
irreversible as in the case of a gas. Indeed, for nearly all experiments which
can be realized in practice, the state $\hat{D}\left(  \tau\right)  $ which
involves these special correlations keeping memory of $\hat{D}\left(
0\right)  $ gives the same predictions as the reduced equilibrium state
$\hat{D}_{\mathrm{R}}$. Nevertheless, the magic sandwich manipulation, which
is rather tricky, transforms back these special correlations of $\hat
{D}\left(  \tau\right)  $ into the initial order restored in $\hat{D}\left(
\tau+\tau^{\prime}+\tau^{\prime\prime}\right)  $.\ The magic sandwich
experiments achieve, for spins, the analogue of the hypothetical reversal of
velocities for a gas; quite remarkably, it is simply the application of some
special uniform time-dependent magnetic fields which overcomes here the
apparent disorder generated by the interactions.

\section{Conclusion}

An analysis of various physical processes where we made use of the information
concept has led us to stress the \textit{relative} nature of entropy,
identified with missing information. Depending on the circumstances,
macroscopic phenomena should be described in terms of different sets of
relevant variables. Different associated entropies are thus introduced, even
for the same system. The entropy of thermodynamics is only one of them.

Dissipation, which is also relative, is measured by the increase of some
relevant entropy. It characterizes the irreversibility of a macroscopic
process, and can be interpreted as an irretrievable leak of information
towards inaccessible degrees of freedom. Since information, as well as
probability, is a concept associated with the knowledge of observers about an
object rather than with the object in itself, thermodynamic notions such as
entropy and dissipation have at the microscopic level a \textit{subjective} aspect.

This subjective character is exemplified by the spin-echo experiments, which
show that \textit{irreversibility itself is relative}. In such circumstances
the loss of memory which accompanies an irreversibility may be overcome and a
hidden order, which keeps track of the initial state, may come into view.

However, in less exotic circumstances, for instance, for systems displaying a
collective motion or for materials that can be described by thermodynamics, it
is possible to select among the huge set of microscopic variables some subset
of relevant variables which obey autonomous equations of motion. The latter
dynamical equations are often identified with phenomenological equations.
Their establishment from microphysics involves approximations, such as the
short-memory approximation, but they can hold with a high precision if there
is a clear-cut separation of time-scales which allows the elimination of
irrelevant variables. The existence of these variables manifests itself
through quasi-instantaneous dissipative effects. Such a reduction of the
description brings in new features, which differ \textit{qualitatively} from
those of microphysics, especially for macroscopic systems: continuity and
extensivity of matter at our scale, irreversibility and non-linearity of the
equations of motion, existence of phase transitions, enormous variety of
behaviours in spite of the simplicity of the elementary constituents and in
spite of the unicity of the microscopic Laws. Due to the change of scale, the
statistical fluctuations of the relevant observables are small compared to
experimental or computational errors. The \textit{very nature of our
description} is thus changed. Although the underlying, more fundamental
description yields predictions having subjective features due to the necessary
use of probabilities, it gives way to a reduced description that no longer
involves the observer. Since the variances are negligible, the physical
quantities do not need to be regarded as random variables. The expectation
values $\left\langle \hat{A}_{i}\right\rangle $ which obey the reduced
dynamics can be interpreted merely as values actually taken in a given system.
Their equations of motion then directly apply to individual objects, without
reference to a statistical ensemble. In spite of the disappearance of
probabilities at the macroscopic scale, the relevant entropy survives as a
quantity measurable indirectly through its variations (\ref{011}) and
characterizing equilibrium, as well known in thermodynamics. Its probabilistic
and subjective origin, its interpretation as missing information are not
apparent; neither are they for the other variables, the expectations $A_{i}$
and the parameters $\gamma_{i}$ of the microscopic probability law. Actually
all these quantities had already been introduced in thermodynamics at the
macroscopic level, but the advent of statistical mechanics gave us a deeper
understanding. Indeed, statistical mechanics is a major tool of reductionism;
its use explains the \textit{emergence} at a larger scale of new properties,
new concepts and even different interpretations of physical statements, such
as here the emergence of objectivity from a microscopic theory based on the
subjective concepts of probability and information. Likewise, treating quantum
measurements as a problem of statistical mechanics (Allahverdyan et al, 2003)
shows how ordinary probabilities emerge from the underlying irreducible
extension of probabilities which accounts for the non-commutation of quantum observables.

Let us finally recall that the identification of entropy with a lack of
information or equivalently with disorder has contributed to the elucidation
of the paradox of \textit{Maxwell's demon}, which gave rise to discussions for
more than one century. An important step was the recognition of the
equivalence between information and negentropy (Brillouin, 1956). The entropy
of a system spontaneously increases; however it may be lowered by some amount,
provided the \textquotedblleft demon\textquotedblright\ or the experimentalist
who acts upon this system makes use of some amount of information at least
equal to the decrease of the entropy. Conversely, in order to acquire some
information, we need to raise the entropy of some system by a quantity at
least equal to the amount of information gained. Information theory thus
enlightens many aspects of statistical physics.

\bigskip

{\small The above text is issued from a talk given at the ESF conference on
philosophical and foundational issues in statistical physics, held in Utrecht
on 28-30 November 2003. I wish to thank Jos Uffink for his invitation to this
conference and for comments.\bigskip}

\section{References}

\noindent Abragam, A. and Goldman, M. (1982). \textit{Nuclear magnetism: order
and disorder} (pp. 45-49). Oxford: Clarendon.

\noindent Allahverdyan, A.E., Balian, R. and Nieuwenhuizen, Th.M. (2003).
Curie-Weiss model of the quantum measurement process. \textit{Europhys. Lett.,
61}, 452-458.

\noindent Balian, R., Alhassid, Y. and Reinhardt, H. (1986). Dissipation in
many-body systems: a geometric approach based on information theory.
\textit{Phys. Reports, 131}, 1-146.

\noindent Balian, R. and Balazs, N. (1987). Equiprobability, inference and
entropy in quantum theory. \textit{Ann. Phys. NY, 179}, 97-144.

\noindent Balian, R. (1989). On the principles of quantum mechanics.
\textit{Amer. J. Phys., 57}, 1019-1027.

\noindent Balian, R. (1991 and 1992). \textit{From microphysics to
macrophysics: methods and applications of statistical physics} (vol. I and
II). Heidelberg: Springer.

\noindent Balian, R. (1999). Incomplete descriptions and relevant entropies.
\textit{Amer. J. Phys., 67}, 1078-1090, and references therein.

\noindent Balian, R. (2004). Entropy, a protean concept. Poincar\'{e} seminar.
\textit{Progress in Math. Phys.} (pp. 119-144). Basel: Birkha\"{u}ser.

\noindent Brillouin, L. (1956). \textit{Science and information theory}. New
York: Academic Press.

\noindent Callen, H.B. (1975). \textit{Thermodynamics}. New York: Wiley.

\noindent Cox, R.T. (1946). Probability, frequency and reasonable expectation.
\textit{Amer. J. Phys., 14}, 1-13.

\noindent de Finetti, B. (1974). \textit{Theory of probability}. New York: Wiley.

\noindent Greenberger, D., Horne, M., Shimony, A. and Zeilinger, A. (1990).
Bell's theorem without inequalities. \textit{Amer. J. Phys., 58}, 1131-1143.

\noindent Jaynes, E.T. (1957).\ Information theory and statistical mechanics.
\textit{Phys. Rev., 106}, 620-630 and \textit{108}, 171-190.

\noindent Lieb, E.H. (1976). The stability of matter. \textit{Rev. Mod. Phys.,
48}, 553-569.

\noindent Mayer, J.E. and Mayer, M.G. (1977). \textit{Statistical mechanics}
(second edition, pp. 145-154). New York: Wiley.

\noindent Ruelle, D. (1969). \textit{Statistical mechanics: rigorous results}.
Reading: Benjamin.

\noindent Shannon, C.E. and Weaver, W. (1949). \textit{The mathematical theory
of communication}. Urbana: University of Illinois Press.

\noindent Thirring, W. (1981). \textit{Quantum mechanics of atoms and
molecules}. New York: Springer.

\noindent Thirring, W. (1983). \textit{Quantum mechanics of large systems}.
New York: Springer.

\noindent Uffink, J. (1995). Can the maximum entropy principle be explained as
a consistency requirement? \textit{Studies in Hist. and Philos. of Mod. Phys.,
26B}, 223-261, and references therein.

\noindent van Kampen, N.G. (1984). The Gibbs paradox. In W.E. Parry (ed.),
\textit{Essays in theoretical physics in honour of Dirk ter Haar} (pp.
303-312). Oxford: Pergamon.

\end{document}